\documentclass{article}

\PassOptionsToPackage{numbers, compress}{natbib}
\usepackage[preprint]{neurips_2024}
\usepackage{fontawesome}

\usepackage{longtable}

\usepackage[utf8]{inputenc} 
\usepackage[T1]{fontenc}    
\usepackage[colorlinks,
            linkcolor=red,
            anchorcolor=blue,
            citecolor=green
            ]{hyperref}
\usepackage{url}            
\usepackage{booktabs}       
\usepackage{amsfonts}       
\usepackage{nicefrac}       
\usepackage{microtype}      
\usepackage{xcolor}         
\usepackage{caption}
\usepackage{subcaption}
\usepackage{natbib}
\usepackage{graphicx}
\usepackage{enumitem}
\usepackage{makecell}
\usepackage{amsmath}
\usepackage{amsthm}
\usepackage{thmtools}
\usepackage{thm-restate}
\usepackage{algorithm}
\usepackage{algorithmic}
\usepackage{bbm}
\usepackage{tabularx}
\usepackage{listings}
\usepackage{xcolor}
\usepackage{multirow}
\usepackage{multicol}
\usepackage{tcolorbox}
\usepackage{tikz}
\usepackage{footnote}
\usepackage{wrapfig}
\usepackage{tikz}

\newsavebox{\leftbox}

\definecolor{tblue}{RGB}{31,119,180}
\definecolor{torange}{RGB}{255,127,14}
\definecolor{tgreen}{RGB}{44,160,44}
\definecolor{tred}{RGB}{214,39,40}
\definecolor{tpurple}{RGB}{148,103,189}
\definecolor{lightblue}{RGB}{173, 216, 230}
\definecolor{lightpink}{RGB}{255, 182, 193}
\definecolor{lightgreen}{RGB}{144, 238, 144}

\usepackage{colortbl}
\usepackage{xcolor}
\usepackage{array}
\usetikzlibrary{fadings}

\newcommand{\hide}[1]{} 

\newcommand{\eg}{\textit{e}.\textit{g}.}

\def\model{DeepCode}

\title{
  \begin{tabular}{@{}c@{}}
    \includegraphics[height=2em]{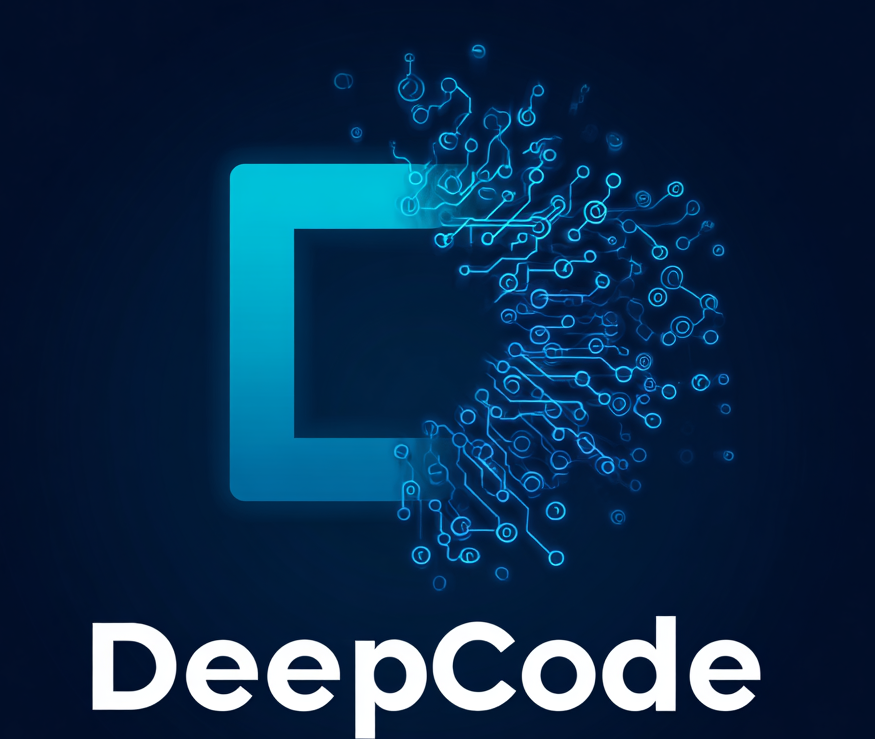} \\
  \end{tabular}
  \hspace{0.5em}
  \begin{tabular}{@{}l@{}}
    DeepCode: Open Agentic Coding
  \end{tabular}
}

\title{\raisebox{-0.8ex}{\includegraphics[height=1.5em]{figs/logo.png}}~DeepCode: Open Agentic Coding}

\lstdefinelanguage{Prompt}{
  moredelim=[s][\color{violet}\bfseries]{`}{`}, 
  moredelim=[s][\color{violet}\bfseries]{<}{>}, 
  morecomment=[s]{<!--}{-->}, 
  morestring=[b]", 
  stringstyle=\color{blue!60}, 
  commentstyle=\color{green}\itshape, 
}
\lstdefinelanguage{Tools}{
literate={[}{{\color{violet}\bfseries[}}1
           {]}{{\color{violet}\bfseries]}}1, 
  morecomment=[s]{<!--}{-->},  
  morestring=[b]",  
  stringstyle=\color{cyan!60},  
  commentstyle=\color{green}\itshape,  
}
\author{
  Zongwei Li\thanks{Equal contribution.}\footnotemark[1] ~~~
  Zhonghang Li\footnotemark[1] ~~~
  Zirui Guo~~~
  Xubin Ren~~~
  Chao Huang\thanks{Chao Huang is the Corresponding Author.} \\
  The University of Hong Kong \\
  \texttt{\{zongwei9888, bjdwh.zzh, larfii1010, xubinrencs, chaohuang75\}@gmail.com}\\
  \faGithub~\textbf{Source Code:} \textcolor{blue}{\url{https://github.com/HKUDS/\model}}
}

\begin{document}

\maketitle

\begin{abstract}
Recent advances in large language models (LLMs) have given rise to powerful coding agents, making it possible for code assistants to evolve into code engineers. However, existing methods still face significant challenges in achieving high-fidelity document-to-codebase synthesis—such as scientific papers to code—primarily due to a fundamental conflict between information overload and the context bottlenecks of LLMs. 
In this work, we introduce \model, a fully autonomous framework that fundamentally addresses this challenge through principled information-flow management. By treating repository synthesis as a channel optimization problem, \model\ seamlessly orchestrates four information operations to maximize task-relevant signals under finite context budgets: source compression via blueprint distillation, structured indexing using stateful code memory, conditional knowledge injection via retrieval-augmented generation, and closed-loop error correction. 
Extensive evaluations on the PaperBench benchmark demonstrate that \model\ achieves state-of-the-art performance, decisively outperforming leading commercial agents such as Cursor and Claude Code, and crucially, surpassing PhD-level human experts from top institutes on key reproduction metrics.
By systematically transforming paper specifications into production-grade implementations comparable to human expert quality, this work establishes new foundations for autonomous scientific reproduction that can accelerate research evaluation and discovery.

\end{abstract}

\begin{figure}[htbp]
    \centering
    \includegraphics[width=0.83\textwidth]{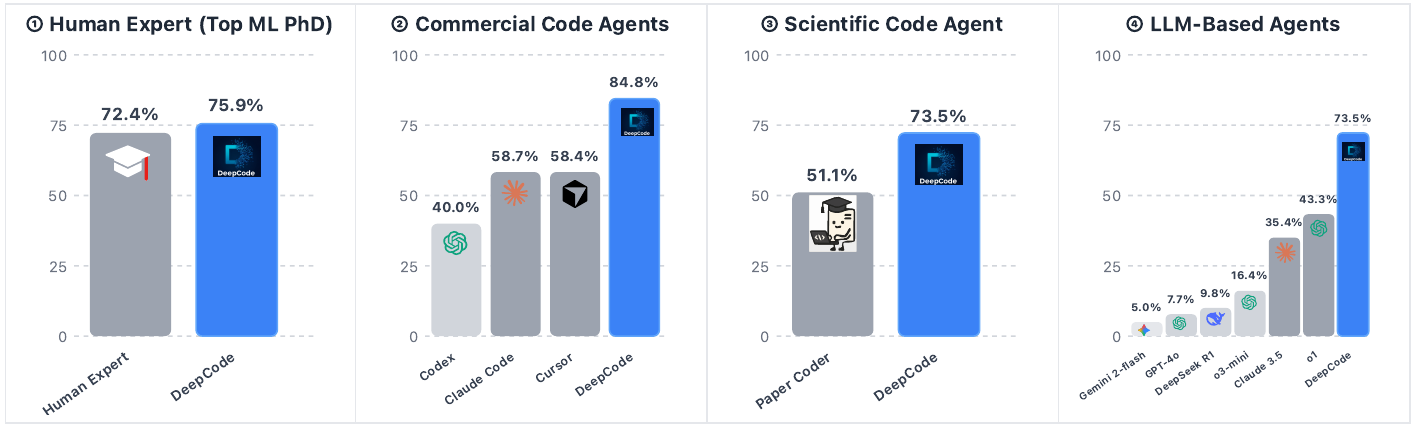}
    \caption{\model\ main results.}
    \label{fig:both}
\end{figure}



\section{Introduction}
\label{sec:intro}

The rapid evolution of Large Language Models (LLMs) has initiated a profound shift in how software is specified, implemented, and maintained~\cite{jiang2024survey,ge2025survey}. AI-assisted coding tools such as Cursor and Codex have already transformed everyday development practice by automating routine implementation tasks and offering intelligent inline suggestions~\cite{peng2023impact,dong2025survey}. Yet these systems remain fundamentally assistive: they operate at the level of code completion, assuming that a human engineer still performs the higher-level tasks of understanding specifications, planning system architecture, and validating behavior. Recent advances in agentic LLM frameworks point toward a more ambitious paradigm---what we term \emph{agentic software engineering}---in which LLM-based agents are expected to plan, orchestrate, and refine entire software projects from high-level natural language or document-level specifications~\cite{wang2025ai,tang2025airesearcher}. In this emerging regime, programming shifts from \emph{writing code} to \emph{writing specifications}, and the central question becomes: \emph{can an artificial coding agent behave as an autonomous engineer that translates rich, informal specifications into comprehensive, robust systems?}


A natural and stringent testbed for this paradigm is \emph{high-fidelity, document-grounded program synthesis}, where a complex scientific paper serves as the sole specification and the goal is to produce a fully executable implementation that faithfully reflects it. Such papers are detailed multimodal specifications, combining informal exposition with equations, pseudo-code, and scattered hyperparameters. In this work, we tackle the highly challenging task of reproducing machine learning papers as complete code repositories. Recent efforts have explored this via LLM-based agents. PaperBench evaluates frontier models on 20 ICML papers, finding the strongest model (o1) with IterativeAgent achieves only 42.4\% replication score, far below 72.4\% for human experts~\cite{starace2025PaperBench}. PaperCoder employs a multi-agent pipeline spanning planning, analysis, and generation, reaching 51.14\% reproduction rate on PaperBench~\cite{seo2025paper2code}. These modest results reveal that current approaches fall well short of reliable, end-to-end replication. We identify four key challenges that underlie this gap:

\begin{figure*}[t]
    \centering
    \includegraphics[width=1\textwidth]{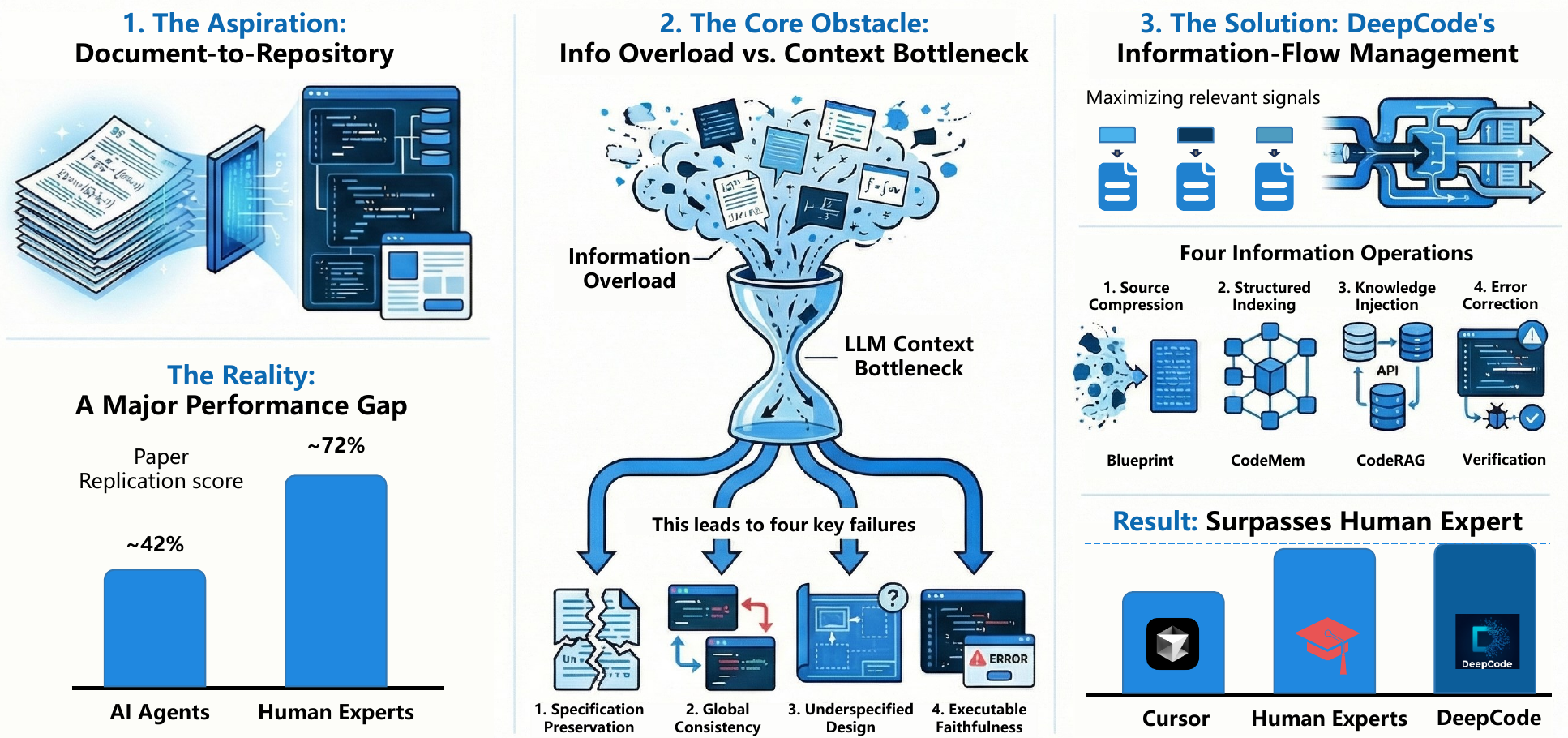}
    \caption{From Challenge to Solution of \model. Left: Current AI agents achieve only a 42\% paper replication score compared to 72\% for human experts, highlighting the limitations of existing agents. Middle: The core challenge stems from information overload conflicting with LLM context limits, causing four key failure modes. Right: \model\ addresses this through four information operations (Blueprint, CodeMem, CodeRAG, Verification), surpassing human expert performance.}
    \vspace{-0.1in}
    \label{fig:intro1}
\end{figure*}

\textbf{(i) Specification Preservation.} Papers describe the target system through scattered, multimodal constraints. Preserving a faithful mapping from this fragmented specification to implementation is inherently difficult.
\textbf{(ii) Global Consistency under Partial Views.} Repositories comprise interdependent modules, but generation proceeds file-by-file under limited context. Maintaining consistency across interfaces, types, and invariants under finite context windows easily leads to broken abstractions.
\textbf{(iii) Completion of Underspecified Designs.} Papers specify only algorithmic cores, leaving implementation details and experimental frameworks implicit. Inferring these consequential but underspecified choices is non-trivial.
\textbf{(iv) Executable Faithfulness.} Faithful reproduction requires executable systems, not just plausible code. Long-horizon generation often yields repositories with subtle logic bugs, dependency conflicts, and fragile pipelines that prevent end-to-end execution.

We argue that fundamentally addressing these challenges requires \emph{principled information-flow management}. We abstract the synthesis process as the transmission of a high-entropy specification—the scientific paper—through a sequence of bandwidth-constrained channels, defined by the LLM's context windows. Naive strategies that simply concatenate raw documents with growing code history induce channel saturation, where redundant tokens mask critical algorithmic constraints, causing the effective Signal-to-Noise Ratio to collapse. Consequently, valid repository generation requires a paradigm shift governed by \emph{contextual information maximization}: at each generation step, the system must actively maximize the density of task-relevant signals while suppressing irrelevant noise. 

Motivated by this perspective, we introduce \textbf{\model}, an open agentic coding framework that fundamentally reimagines repository-level synthesis as a problem of \emph{hierarchical information-flow management}. Rather than treating synthesis as a monolithic process, \model\ systematically addresses the doc-to-repos challenges by instantiating the proposed paradigm through four orchestrated information operations: (1) \emph{source compression}, which distills unstructured multi-modal specifications into a precise structural blueprint to maximize signal density; (2) \emph{structured indexing}, which abstracts the evolving repository state into concise memory entries to maintain global consistency without context saturation; (3) \emph{conditional knowledge injection}, which leverages retrieval-augmented generation to bridge implicit specification gaps with standard implementation patterns; and (4) \emph{error correction}, which utilizes closed-loop verification to transform execution feedback into corrective signals for rectifying transmission errors. Our contributions are threefold:

\begin{itemize}[leftmargin=*]
    \item We characterize the task of high-fidelity document-to-repository synthesis through an information-theoretic lens, identifying the central conflict as an \emph{information-overload vs. context-bottleneck} conflict. From this perspective, we propose an information-theoretic design principle: effective agentic coding systems must explicitly structure, route, and compress information to maximize task-relevant signal under finite context budgets.
    \item We instantiate this principle in \model, a systematic framework that orchestrates four strategic information operations: blueprint distillation, stateful memory management, conditional knowledge injection, and closed-loop verification. By dynamically optimizing the signal-to-noise ratio within the context window, \model\ effectively resolves the challenges of long-range specification preservation, cross-file consistency, and implicit knowledge gaps in complex generation tasks.
   \item Extensive evaluations on the PaperBench benchmark demonstrate that \model\ achieves state-of-the-art performance, decisively \textbf{outperforming leading commercial agents} (\eg~Cursor, Claude Code, Codex) and, notably, \textbf{surpassing human expert performance} on key reproduction metrics. Furthermore, our analysis reveals that principled information-flow management yields significantly larger performance gains than merely scaling model size or context length, offering a pivotal direction for the future design of autonomous software engineers.
\end{itemize}

\section{Preliminary}
\label{sec:preliminary}

\subsection{Task Definition}

The primary objective of this work is to develop a system for \textit{high-fidelity program synthesis}. We formalize this as the process of learning a mapping function, $\mathcal{F}_{gen}$, which transforms a specification document, $\mathcal{D}$, into a complete and executable code repository, $\mathcal{P}$. The core function is defined as:
\begin{equation}
    \mathcal{F}_{gen}: \mathbb{D} \rightarrow \mathbb{P}
\end{equation}
where $\mathbb{D}$ represents the space of specification documents and $\mathbb{P}$ represents the space of valid code repositories. Such that for a given input document $\mathcal{D} \in \mathbb{D}$, the output is a program repository $\mathcal{P} = \mathcal{F}_{gen}(\mathcal{D})$. We address two primary manifestations of this task:

\vspace{-0.05in}
\begin{itemize}[leftmargin=*]
    \item \textbf{Scientific Paper Reproduction:} Given a scientific paper from domains such as machine learning or computer sciences as the source document $\mathcal{D}$, the system should generate the full source code $\mathcal{P}$ required to replicate the paper's key experiments and results.
    \item \textbf{Software System Generation:} Given a comprehensive technical design document or a concise natural language requirement for a software application (e.g., specifying UI, backend APIs, and database schema) as $\mathcal{D}$, the system should generate the corresponding multi-component software repository $\mathcal{P}$, including frontend, backend, and configuration files.
\end{itemize}
\vspace{-0.05in}


\textbf{Input: Source Document $\mathcal{D}$.}
The source document $\mathcal{D}$ is represented as a sequence of multi-modal elements, $\mathcal{D} = (d_1, d_2, \dots, d_L)$, where each element $d_i$ can be a block of text, a mathematical equation, a table, a figure, or a snippet of pseudocode. The length $L$ of this sequence is typically large, posing significant challenges for models with finite context windows.

\textbf{Output: Code Repository $\mathcal{P}$.}
The target output $\mathcal{P}$ is not a single file but a structured repository. We define it as a tuple:
\begin{equation}
    \mathcal{P} = (\mathcal{T}, \mathcal{C}, \mathcal{M})
\end{equation}

Here, $\mathcal{T}$ represents the directory structure that organizes the files in $\mathcal{C}$. $\mathcal{C} = \{c_1, c_2, \dots, c_N\}$ is a set of $N$ source code files. The generation of a coherent set $\mathcal{C}$ where files correctly interact (e.g., via imports and function calls) is a non-trivial problem of ensuring cross-file consistency. $\mathcal{M}$ is the dependency manifest (\eg~\texttt{requirements.txt},  \texttt{package.json}, \texttt{README.md} file) specifying all external libraries required to run the code.

\subsection{Objectives}

An ideal synthesis function $\mathcal{F}_{gen}$ must generate a repository $\mathcal{P}^*$ that optimizes a composite scoring function. Under our paradigm of \emph{principled information-flow management}, this optimization is framed as maximizing the effective signal-to-noise ratio across the synthesis channel. The optimal output is defined as:
\begin{equation}
    \mathcal{P}^* = \arg\max_{\mathcal{P} \in \mathbb{P}} \text{Score}(\mathcal{P} | \mathcal{D})
\end{equation}

To overcome the conflict between information overload and finite context bandwidth, the scoring function decomposes into four distinct objectives, each corresponding to an information operation:

\vspace{-0.05in}
\begin{itemize}[leftmargin=*]
    \item \textbf{Specification Preservation:} The repository must faithfully implement the rigid algorithmic constraints hidden within the multimodal source document. The objective is to maximize signal density by extracting precise blueprints from the unstructured input noise.
    
    \item \textbf{Global Structural Consistency:} The generated modules must maintain strict interface compatibility and type coherence. The objective is to maintain state consistency without context saturation, achieved by indexing the evolving codebase into compact, retrievable summaries.
    
    \item \textbf{Domain Knowledge Grounding:} The system must bridge the gap between abstract academic descriptions and concrete engineering implementations. The objective is to resolve underspecified designs by conditionally injecting standard libraries and patterns from external knowledge bases.

    \item \textbf{Functional Executability:} The final repository must be robust and runnable. The objective is to minimize transmission errors (bugs) by treating runtime execution feedback as a corrective signal to iteratively refine the generated code.
\end{itemize}
\vspace{-0.05in}

Our framework is designed to satisfy these objectives by explicitly routing and compressing information, enabling high-fidelity repository generation under strict context window constraints.
\section{The \model\ Framework}
\label{sec:methodology}


We introduce \model, a multi-stage framework designed to instantiate the principle of principled information-flow management for repository-level synthesis. To solve the optimization problem, \model\ decomposes the generation process into three orchestrated phases, each serving a distinct information-processing role to maximize the effective signal-to-noise ratio. The process initiates with \textbf{(1) Blueprint Generation}, where a planning agent acts as a source compression mechanism, distilling the high-entropy source document $\mathcal{D}$ into a structured, high-signal implementation blueprint to extract critical constraints while filtering narrative noise. Guided by this blueprint, the subsequent \textbf{(2) Code Generation} phase synthesizes source files while preventing channel saturation through two integrated mechanisms: a stateful Code Memory (CodeMem) that performs structured indexing of the evolving codebase to maintain cross-file consistency, and a CodeRAG system that performs conditional knowledge injection to bridge implicit domain gaps with standard implementation patterns. Finally, the framework concludes with \textbf{(3) Automated Verification}, a closed-loop error correction phase where a validation agent treats runtime execution feedback as corrective signals to identify and rectify transmission errors, ensuring the functional correctness of the final output.

\subsection{Phase 1: Blueprint Generation}


The primary goal of the first phase is to perform source compression: distilling the unstructured, lengthy content of a source document (\eg~a scientific paper) into a structured, machine-readable implementation blueprint. This distillation process directly mitigates the challenges of information overload by transforming the raw input $\mathcal{D}$ into a high-density signal format. The process begins with a crucial preprocessing step: hierarchical content segmentation.

\begin{figure*}[t]
    \centering
    \includegraphics[width=1\textwidth]{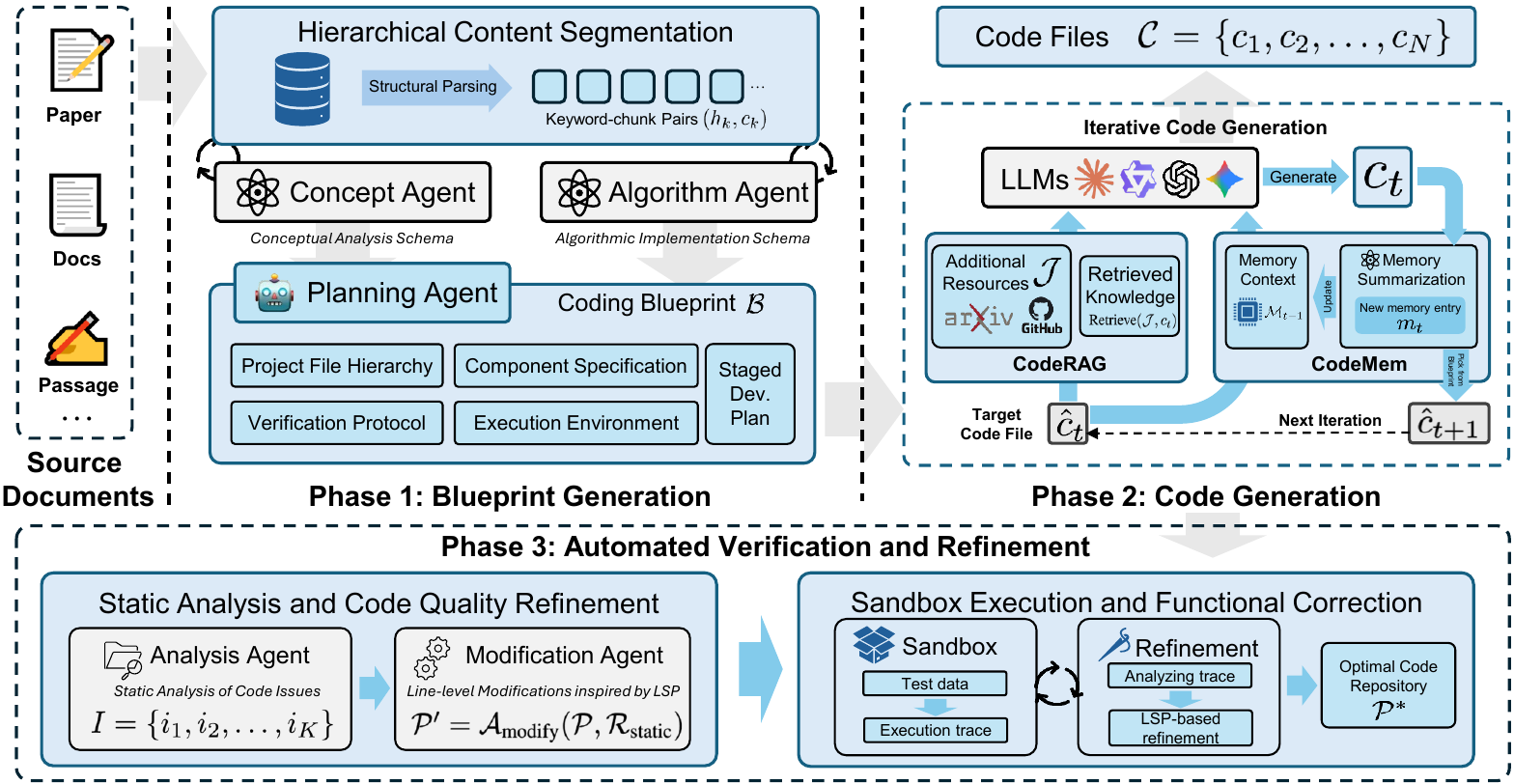}
    \caption{The overall framework of DeepCode.}
    \vspace{-0.1in}
    \label{fig:intro1}
\end{figure*}

\subsubsection{Hierarchical Content Segmentation}

Instead of feeding the entire document $\mathcal{D}$ into an LLM, we first parse it into a structured representation that facilitates targeted information access. We introduce a \textbf{hierarchical content index}, which leverages the inherent structure of academic papers and technical documents. The process is:

\vspace{-0.05in}
\begin{enumerate}[leftmargin=*]
    \item \textbf{Structural Parsing:} The source document $\mathcal{D}$ is parsed to identify its hierarchical structure based on explicit delimiters like section and subsection titles (\eg~"3. Methodology", "3.1. Model Architecture"). This divides the document into a set of content chunks $S = \{s_1, s_2, \dots, s_K\}$.

    \item \textbf{Keyword-Chunk Association:} Each chunk $s_k$ is stored as a key-value pair $(h_k, c_k)$, where the heading $h_k$ serves as a natural, high-level semantic keyword, and $c_k$ is the corresponding raw text content of that section.
\end{enumerate}
\vspace{-0.05in}

This indexed structure effectively transforms the problem from one of long-context comprehension to a series of more manageable, on-demand retrievals. An agent no longer needs to process the entire document at once. Instead, it can query the index using semantic keywords (\eg~requesting the content associated with "Model Architecture") to fetch only the most relevant context for its current task. This approach drastically reduces the token load for any single operation and allows the model to focus its limited context window on the most pertinent information, thereby solving the problem of context overload and information forgetting. This structured representation serves as the foundational input for the specialized agents that perform the detailed analysis in the subsequent steps.

\subsubsection{Multi-Agent Specification Analysis}

Following the hierarchical segmentation, we employ a specialized multi-agent system to conduct a deep and structured analysis of the document's content. This approach decomposes the complex comprehension task into two parallel tracks, executed by a \textbf{Concept Agent} and an \textbf{Algorithm Agent}. Each agent is equipped with a specific prompt and interacts with the indexed document to extract complementary layers of information, ensuring a comprehensive understanding without processing the entire document simultaneously.

\textbf{Concept Agent: High-Level Structural and Conceptual Mapping.}
The Concept Agent is tasked with building a holistic, high-level understanding of the document. Its primary objective is to map the paper's entire conceptual structure, identify its core scientific contributions, and outline the necessary components for a successful experimental reproduction. Operating on the indexed document, the agent is instructed to use a segmented reading strategy, querying the index with semantically broad keywords (\eg~``introduction'', ``method''). This allows it to assemble a comprehensive overview by strategically fetching relevant sections. The output of this agent is a structured \textit{Conceptual Analysis Schema}. This schema comprises a detailed paper structure map, a method decomposition map outlining the system’s core functional components, an implementation map aligning claims with code requirements, and a reproduction roadmap specifying the criteria for success. Collectively, these elements translate the paper’s narrative into a structured project plan.

\textbf{Algorithm Agent: Low-Level Technical Detail Extraction.}
Complementing the conceptual overview, the Algorithm Agent is responsible for the meticulous extraction of every low-level technical detail required for an exact implementation. It's designed to perform an exhaustive search for all algorithms, mathematical formulations, model architectures, training procedures, and hyperparameters. Moreover, it can leverage online search capabilities to retrieve relevant algorithm implementations from the web as references. Like the Concept Agent, it leverages the segmented reading strategy but uses a distinct set of highly specific keywords (\eg~``algorithm'', ``hyperparameter'') to perform targeted queries on the most technically dense sections of the document. The agent's output is a granular \textit{Algorithmic Implementation Schema}. This schema captures verbatim pseudocode from algorithm boxes, exact mathematical equations and their variables, detailed layer-by-layer network architectures, and a comprehensive list of all hyperparameters with references to their locations in the paper. This schema serves as a precise, unambiguous technical specification, designed to leave no detail to interpretation during the code generation phase.

\subsubsection{Synthesizing the Implementation Blueprint}

The analytical outputs from the Concept and Algorithm agents are then synthesized by the \textbf{Code Planning Agent} into a single, holistic implementation blueprint. This agent's critical function is to orchestrate the high-level conceptual framework with the low-level technical specifications, performing a final disambiguation and grounding step. It reconciles the architectural overview with the granular implementation details, ensuring that every abstract component is directly linked to a precise technical specification. Should any inconsistencies arise, the agent is authorized to perform targeted queries on the indexed document to resolve them.
The final \textbf{Implementation Blueprint} $\mathcal{B}$ is a structured intermediate representation designed to be a self-contained, unambiguous specification for code generation. This blueprint is organized into the following canonical sections:

\vspace{-0.05in}
\begin{itemize}[leftmargin=*]
    \item \textbf{Project File Hierarchy:} A prioritized project file structure that dictates the logical organization of the codebase and the implementation order of its modules.

    \item \textbf{Component Specification:} A granular specification for every module, class, and function, explicitly mapping each to its corresponding algorithmic pseudocode and mathematical formulation.

    \item \textbf{Verification Protocol:} A formal plan for validating the final implementation. It defines the experimental setup, specifies the target metrics from the source document, and establishes the success criteria for reproduction.

    \item \textbf{Execution Environment:} A complete specification of all software dependencies, library versions, and requisite hardware configurations needed to compile and run the code.

    \item \textbf{Staged Development Plan:} A phased implementation roadmap that defines the build order of components and integrates staged verification checks to ensure modular correctness.
\end{itemize}
\vspace{-0.05in}

By consolidating all distilled information into this canonical blueprint, the Code Planning Agent concludes the specification distillation phase. This artifact serves as the definitive "source of truth" for the subsequent code generation phase, effectively resolving the long-context challenge by providing a dense, structured, and actionable input that obviates any need for the coding agents to interact with the original, lengthy document.

\subsection{Phase 2: Code Generation}
\label{sec:code_generation}


Upon generating the high-signal blueprint, the second phase synthesizes the code repository. This phase maximizes the density of relevant context while preventing channel saturation caused by the accumulation of raw code history. A naive iterative approach, which appends previously generated code to the prompt, leads to a collapse in the signal-to-noise ratio and induces hallucinations. To overcome this, we propose a dual-mechanism strategy for efficient information routing: (1) a stateful \textbf{CodeMem} that performs structured indexing of the evolving repository to maintain internal structural cohesion without context bloat, and (2) a \textbf{CodeRAG} system that performs conditional knowledge injection, grounding the implementation in external patterns to bridge implicit knowledge gaps.

\subsubsection{Stateful Generation with CodeMem}

The core of our generation process is the Code Memory mechanism, a strategy designed to maintain a compressed, structured representation of the repository's state, thereby ensuring cross-file consistency without suffering from prohibitive context lengths. Instead of passing the full source code of previously implemented files to the generative agent, we iteratively build and query a structured memory bank, $\mathcal{M}$.

Let the set of all files to be implemented, as defined by Sec.~\ref{sec:preliminary}, be $\mathcal{C} = \{c_1, c_2, \dots, c_N\}$. The generation process is an iterative loop over $t=1, \dots, N$. At each step $t$, we maintain the set of implemented files, $\mathcal{C}_{t-1}$, and the set of unimplemented files, $\mathcal{U}_{t-1}$. The process for generating the target file for the current step, $\hat{c}_t$, is as follows:

\vspace{-0.05in}
\begin{enumerate}[leftmargin=*]
    \item \textbf{Context Formulation.} The generation context for the current step, $\mathcal{X}_t$, is constructed not from raw source code, but from the static implementation blueprint $\mathcal{B}$ and a dynamically selected subset of the Code Memory, $\mathcal{M}_{t-1}$. The agent first identifies which previously implemented files are relevant to the current target file $\hat{c}_t$ (where $\hat{c}_t$ denotes the blank code file to be generated, and $c_t$ denotes the resulting generated code file). It then retrieves only their corresponding summaries from the memory bank:
    \begin{equation}
        \mathcal{X}_t = \left( \mathcal{B}, \text{SelectRelevantMemory}(\mathcal{M}_{t-1}, \hat{c}_t) \right)
    \end{equation}
    where $\text{SelectRelevantMemory}$ is a function that queries $\mathcal{M}_{t-1}$ to fetch only the essential summaries of dependencies.

    \item \textbf{Code Generation.} The coding agent, represented by the LLM function $\mathcal{L}$, synthesizes the source code for the target file based on the curated context:
    \begin{equation}
        c_t = \mathcal{L}(\mathcal{X}_t)
    \end{equation}

    \item \textbf{Memory Update.} After generating the code $c_t$, the system clears the generation context. A specialized summarization agent, $\mathcal{S}$, is then invoked. This agent analyzes the newly generated source code $c_t$ to extract its structural essence and create a new memory entry, $m_t$. The Code Memory is then updated:
    \begin{equation}
        \mathcal{M}_t = \mathcal{M}_{t-1} \cup \{m_t\}
    \end{equation}
\end{enumerate}
\vspace{-0.05in}

The summarization agent $\mathcal{S}$ distills the code into a structured format that captures all information necessary for inter-module communication. Each memory entry $m_t$ is a structured object containing:

\vspace{-0.05in}
\begin{itemize}[leftmargin=*]
    \item \textbf{Core Purpose ($\mathcal{P}_t$):} A concise, natural language summary of the file's primary responsibility and role within the repository.

    \item \textbf{Public Interface ($\mathcal{I}_t$):} A formal description of all externally accessible classes, functions, and constants, including their signatures and purposes (e.g., Class(params): methods).

    \item \textbf{Dependency Edges ($\mathcal{E}_t$):} A comprehensive map of the file's position within the project's dependency graph. This structured entry specifies both \textbf{afferent couplings} (internal dependencies), detailing the specific imports from other project modules and external packages, and predicted \textbf{efferent couplings} (external dependencies), identifying which unimplemented modules are expected to consume this file's public interface. 

    \item \textbf{Next Implementation Target ($\hat{c}_{t+1}$):} A decision on the next file to be implemented, based on the blueprint, dependency graph and the current state. Note that, to avoid introducing noise into the memory, this information is separated from $m_t$ and provided independently as part of $\mathcal{L}$ input.
\end{itemize}
\vspace{-0.05in}

This mechanism effectively decouples the context size from the repository size. The context provided to the agent at any step $t$ remains compact, containing only the high-level blueprint and the highly compressed summaries of relevant, already-implemented files. This stateful, summary-based approach allows our system to maintain global consistency and logical cohesion across a large number of files, directly solving the long-context and cross-file consistency challenges.

\subsubsection{Knowledge Grounding with CodeRAG}

While the Code Memory mechanism ensures internal consistency, it does not address the challenges of model hallucination or the omission of implicit domain knowledge. To mitigate these issues, we introduce a retrieval-augmented generation framework, \textbf{CodeRAG}, which grounds the synthesis process in a pre-indexed corpus of relevant, high-quality code repositories. This process is divided into two stages: an indexing phase and an adaptive retrieval phase during code generation.

\textbf{Repository Indexing.}
The goal of this phase is to analyze a set of relevant source code repositories, $\mathcal{R} = \{R_1, R_2, \dots, R_K\}$, and build a structured, queryable index, $\mathcal{J}$. The process, modeled by $\mathcal{I}_{\text{index}}: \mathcal{R} \times \mathcal{B} \rightarrow \mathcal{J}$, consists of the following steps:

\vspace{-0.05in}
\begin{enumerate}[leftmargin=*]
    \item \textbf{Relevance Filtering:} For each repository $R_k \in \mathcal{R}$, we perform an initial LLM-based filtering to identify a subset of source files, $\mathcal{C}'_k \subset R_k$, that are most relevant to the target project structure defined in the implementation blueprint $\mathcal{B}$. In this context, $\mathcal{R}$ can denote either the corresponding repository cited in the references of the target paper or other relevant repositories identified through online search. This focuses computational resources on the most promising assets.

    \item \textbf{Code Understanding:} Each relevant source file $c_s' \in \mathcal{C}'_k$ is independently analyzed to create a structured summary, analogous to the memory entries described previously. This summary captures the file's purpose, key concepts, and public interfaces.

    \item \textbf{Relationship Mapping:} The core of the indexing process is to establish explicit links between the analyzed source files and the target files in our blueprint. For each source file summary, an agent maps it to one or more target files in $\mathcal{B}$, generating a set of relationship tuples.
\end{enumerate}
\vspace{-0.05in}

The final output index $\mathcal{J}$ is a structured knowledge base containing a collection of relationship tuples. Each tuple is defined as $(c_s', \hat{c_t}, \tau, \sigma, \gamma)$. Here, $c_s'$ is a file in the source repository and $\hat{c_t}$ is the corresponding target file in the blueprint's structure.
$\mathbf{\tau}$ denotes the relationship type, indicating the nature of the potential contribution, while $\mathbf{\sigma}$ is a confidence score representing the strength of the mapping.
$\mathbf{\gamma}$ is a set of actionable context, such as helpful code snippets, usage suggestions, and implementation patterns.

\textbf{Adaptive Retrieval.}
During the iterative code generation phase, our framework will optionally query the CodeRAG index $\mathcal{J}$ to augment its context. At each generation step $t$ for a target file $\hat{c_t}$, the agent makes an adaptive decision on whether to retrieve external knowledge. This decision is modeled by a binary function $\delta$:
\begin{equation}
    r_t = \delta(\mathcal{X}_t, \hat{c_t})
\end{equation}
where flag $r_t \in \{0, 1\}$ and $\mathcal{X}_t$ is the standard context containing the blueprint and relevant code memory. The decision is based on the complexity of the target file and the level of detail available in the blueprint.
If $r_t = 1$, the agent queries the index $\mathcal{J}$ to find the most relevant relationship tuples for $\hat{c_t}$. The retrieved context $\gamma$ from the highest-confidence relationship is used to create an augmented context, $\mathcal{X}'_t$:
\begin{equation}
    \mathcal{X}'_t = \mathcal{X}_t \cup \{\text{Retrieve}(\mathcal{J}, \hat{c_t})\}
\end{equation}
The final code is then generated using this enriched context: $c_t = \mathcal{L}(\mathcal{X}'_t)$. By dynamically incorporating proven implementation patterns from existing repositories, CodeRAG significantly reduces the likelihood of generating erroneous or suboptimal code, thus bridging the knowledge gap for the generative agent.

\subsection{Phase 3: Automated Verification and Refinement}
\label{sec:verification}


The final phase serves as an error correction mechanism to ensure the functional faithfulness of the synthesized repository $\mathcal{P}$. Recognizing that purely generative processes are prone to transmission errors—manifesting as logic bugs, invalid dependencies, or dead code—this phase establishes a crucial closed-loop feedback system absent in standard models. By treating execution outcomes as corrective signals, the framework systematically identifies and rectifies defects through two sequential stages: (1) a static analysis pass to ensure structural integrity and code quality, and (2) a dynamic execution pass within a sandboxed environment to enforce functional correctness.

\subsubsection{Static Analysis and Code Quality Refinement}

The first stage addresses issues that can be detected without executing the code. This process is orchestrated by a dedicated Analysis Agent and a Modification Agent.

\textbf{Static Analysis.}
An Analysis Agent, denoted by the function $\mathcal{A}_{\text{static}}$, inspects the generated repository $\mathcal{P}$ against the implementation blueprint $\mathcal{B}$. It produces a structured static analysis report, $\mathcal{R}_{\text{static}}$, which identifies a set of issues. This process can be formalized as: $\mathcal{R}_{\text{static}} = \mathcal{A}_{\text{static}}(\mathcal{P}, \mathcal{B})$.

The identified issues $I = \{i_1, i_2, \dots, i_K\}$ fall into two categories:
i) \emph{Structural Discrepancies:} This includes integrity violations such as missing files specified in the blueprint or empty (zero-byte) source files that were not correctly generated.
ii) \emph{Code Quality Deficiencies:} The agent leverages an LLM to perform a quality assessment of each source file, assigning a quality score, $q(c_i)$, and flagging sections with poor style, complexity, or maintainability.

\textbf{Code Refinement.}
The report $\mathcal{R}_{\text{static}}$ is then passed to a Modification Agent, $\mathcal{A}_{\text{modify}}$. This agent iterates through each issue $i_k \in I$ and applies a targeted fix. To perform precise, line-level modifications without rewriting entire files, the agent utilizes a programmatic interface inspired by the Language Server Protocol (LSP). We model this refinement operation as a function $\Phi_{\text{LSP}}$ that takes a file $c_i$ and a modification instruction from the report, producing a corrected file $c'_i$. The overall process yields a statically refined repository $\mathcal{P}'$ as: $\mathcal{P}' = \mathcal{A}_{\text{modify}}(\mathcal{P}, \mathcal{R}_{\text{static}})$.

\subsubsection{Sandbox Execution and Functional Correction}

After static refinement, the repository $\mathcal{P}'$ undergoes dynamic testing in a secure, isolated sandbox environment to ensure it runs as intended.

\textbf{Environment Verification and Setup.}
A Sandbox Agent, $\mathcal{A}_{\text{sandbox}}$, first validates the environment setup instructions (e.g., in \texttt{README.md}) against the dependencies specified in the blueprint $\mathcal{B}$. Any discrepancies are corrected. The agent then automatically provisions the specified environment and installs all dependencies.

\textbf{Iterative Execution and Correction.}
The agent then attempts to execute the main entry points of the repository, using automatically generated test data and test files designed to exercise the core algorithms and functions. The execution process, $\mathcal{E}_{\text{sandbox}}$, takes the repository $\mathcal{P}'_j$ at iteration $j$ (initially $\mathcal{P}'_0 = \mathcal{P}'$) and produces an execution trace, $\mathcal{T}_j$, containing all outputs and error messages.
\begin{equation}
    \mathcal{T}_j = \mathcal{E}_{\text{sandbox}}(\mathcal{P}'_j)
\end{equation}
This initiates an iterative refinement loop. If the trace $\mathcal{T}_j$ contains errors ($\mathcal{T}_j^{\text{error}} \neq \emptyset$), the Sandbox Agent analyzes the error messages to identify the likely faulty files and the nature of the bug. It then generates a modification instruction and invokes the LSP-based refinement function $\Phi_{\text{LSP}}$ to patch the code, producing the repository for the next iteration, $\mathcal{P}'_{j+1}$. This loop continues until the execution is successful or a maximum number of iterations is reached.
\begin{equation}
    \mathcal{P}'_{j+1} = \Phi_{\text{LSP}}(\mathcal{P}'_j, \mathcal{T}_j^{\text{error}})
\end{equation}
The final verified output of our entire framework is the repository $\mathcal{P}^* = \mathcal{P}'_J$, where $J$ is the terminal iteration of the refinement loop. This multi-stage verification and correction process ensures that the synthesized code is not only structurally sound but also functionally correct and conformant to the original specification.
\section{Experiments}
\label{sec:eval}
In this section, we evaluate the effectiveness of the proposed \model\ framework by addressing the following 3 research questions:
\textbf{RQ1:} How does \model\ perform compared to existing agent frameworks?
\textbf{RQ2:} How does the choice of different LLMs affect the performance of \model?
\textbf{RQ3:} What is the contribution of each module within the \model\ architecture?

\subsection{Experiments Settings}
\textbf{Datasets.}
To evaluate \model's capabilities in code comprehension and generation, particularly for automated vulnerability detection, we employ \textbf{PaperBench Code-Dev}, an innovative benchmark created by OpenAI \cite{starace2025PaperBench}. PaperBench Code-Dev assesses AI models' ability to independently reproduce leading ML research from major conferences like ICML 2024, focusing on 20 significant papers. Models are required to generate all necessary code from scratch, using only the research papers as references, without accessing existing codebases from the original authors. These tasks are performed in a virtual machine environment, with the goal of building a functional codebase, replicating experiments, and creating a \texttt{reproduce.sh} script for execution. Each paper is accompanied by a detailed evaluation rubric approved by the authors, which breaks down the reproduction task into 8,316 specific, gradable components, meticulously assessed using a hierarchical weighting scheme and SimpleJudge, a sophisticated automated judge powered by OpenAI's o3-mini model. This benchmark is rigorously crafted to challenge AI with tasks requiring advanced natural language understanding, algorithmic reasoning, and the ability to generate reliable code from abstract descriptions, all of which are crucial skills for automating vulnerability detection effectively.

\textbf{Baselines.} 
In order to evaluate the effectiveness of the proposed framework, we include a range of baseline methods for comparison. These baselines fall into four distinct categories:

\textbf{(1) LLM Agents.} We compare against results reported in~\cite{starace2025PaperBench} for several state-of-the-art language models using two agent scaffolding approaches: (1) \textit{BasicAgent}, a simple tool-use loop based on Inspect AI's basic agent that allows models to terminate early, and (2) \textit{IterativeAgent}, which forces models to use their full allocated time and employs prompts designed to encourage incremental, piecemeal progress. All agents run in Ubuntu 24.04 Docker containers with access to a single A10 GPU, the internet, and standard development tools including bash, Python, web browsing, and file reading capabilities~\cite{starace2025PaperBench}. The baseline models include GPT-4o, o1, o3-mini, DeepSeek-R1, Claude 3.5 Sonnet, and Gemini 2.0 Flash, with most experiments using a 12-hour time limit (extended to 36 hours for select o1 runs).

\textbf{(2) Scientific Code Agents.} \textit{PaperCoder}~\cite{seo2025paper2code}. PaperCoder (also referred to as Paper2Code) is a multi-agent LLM framework that transforms machine learning papers into executable code repositories via a three-stage pipeline: planning, which constructs implementation roadmaps, system architecture diagrams, and file dependencies; analysis, which extracts file-level implementation details; and generation, which produces modular code in dependency order.

\textbf{(3) Commercial Code Agents.} We compare against three state-of-the-art commercial code agents that provide AI-powered development assistance through different interfaces and capabilities:

\vspace{-0.05in}
\begin{itemize}[leftmargin=*]
\item  \textit{Cursor} (Version 1.7.52) is an AI-assisted integrated development environment built as a fork of Visual Studio Code with additional AI features. Cursor allows developers to choose between cutting-edge LLMs and provides codebase embedding models that give agents deep understanding and recall~\cite{cursor2025}. In our experiments, Cursor uses Claude Sonnet 4.5-thinking as the underlying model.

\item \textit{Claude Code} (Version 2.0.22) is Anthropic's agentic coding tool that lives in the terminal and helps developers turn ideas into code. Claude Code maintains awareness of the entire project structure, can find up-to-date information from the web, and with MCP can pull from external data sources like Google Drive, Figma, and Slack. It can directly edit files, run commands, create commits, and use MCP to read design docs or update tickets~\cite{claudecode2025}. Our evaluation uses Claude Sonnet 4.5-thinking.

\item \textit{Codex} (Version codex-cli 0.47.0) is OpenAI's coding agent that runs locally from the terminal and can read, modify, and run code on the user's machine. Codex is optimized for use with GPT-5-Codex for agentic coding, with configurable reasoning levels from medium to high for complex tasks. In auto approval mode, Codex can read files, make edits, and run commands in the working directory automatically~\cite{codex2025}. We configure Codex with GPT-5 Codex-high.
\end{itemize}
\vspace{-0.05in}

\textbf{(4) Human Experts.} The human baseline~\cite{starace2025PaperBench} consists of 8 ML PhD students and graduates from top institutions (\eg~Berkeley, Cambridge, Carnegie Mellon) who worked part-time over a four-week window on a 3-paper subset (all-in-one, fre, stay-on-topic). Participants had similar computational resources (A10 GPU) and could use AI coding assistants like ChatGPT and GitHub Copilot. The best-of-3 human attempts (Best@3) represent expert-level performance on this subset.

\textbf{Experimental Setup.} To evaluate \model's efficacy in high-fidelity repository synthesis, we adopt a rigorous framework under realistic constraints. The setup combines a secure execution environment and the PaperBench protocol for fair, reproducible, detailed comparisons across baselines.

\textbf{(1) Implementation Environment.} All experiments are conducted within an Ubuntu 22.04 LTS-based sandboxed environment. This infrastructure is provisioned with a standard Python development stack and essential dependencies. \model\ is configured to operate within this isolated space, retaining privileges for file system manipulation, shell command execution, and internet access, thereby simulating a standard software research and development workflow.

\textbf{(2) Task Execution.} \model\ accepts the target paper in both PDF and Markdown formats, along with any supplementary addenda, as primary inputs. To ensure that generated solutions stem from algorithmic reasoning rather than retrieval, a source code blacklist is enforced during execution. This protocol precludes access to the authors' original repositories and known third-party implementations during web browsing. With input parameters defined and the search space constrained, \model\ initiates its autonomous workflow for code generation and debugging.

\textbf{(3) Grading Methodology.}
Assessment of the generated code follows the PaperBench Code-Dev protocol, which focuses on structural and functional correctness and does not include post-submission reproduction. Grading is carried out by SimpleJudge, an automated system based on OpenAI's o3-mini, which performs static analysis of the submitted repository against a set of fine-grained, hierarchical criteria co-developed with the authors of the source paper. The judging logic is restricted to the ``Code Development'' leaf nodes of this rubric and examines core aspects of software quality, including static correctness (syntax validity and compliance with language standards), dependency validity (completeness and correctness of dependency specifications such as \texttt{requirements.txt}), project structure (coherent and consistent organization of files and directories), and algorithmic fidelity (faithful implementation of the algorithms and interfaces described in the original paper). This procedure is designed to align the evaluation with the central technical contributions of the work.

\textbf{(4) Evaluation Metrics and Protocol.} Our primary evaluation metric is the Replication Score, which quantifies the proficiency of \model\ in translating theoretical concepts into a functional codebase. The score for a single replication trial is derived from the hierarchical rubric through a bottom-up aggregation process. \textbf{(i) Leaf node scoring:} SimpleJudge first evaluates each leaf node criterion on a binary basis, assigning a score of 1 for ``pass'' (compliance) and 0 for ``fail'' (non-compliance). \textbf{(ii) Score aggregation:} The score for any parent node is then computed as the weighted average of the scores of its immediate children. The weights, predetermined during the rubric design, reflect the relative importance of each sub-task. \textbf{(iii) Final score derivation:} This recursive aggregation continues up the hierarchy until a single score is obtained for the root node, which serves as the Replication Score for that trial.

To account for the stochasticity inherent in code generation, we adopt a strict evaluation protocol. For each target paper, three independent replication trials are performed, and each resulting repository is scored separately by SimpleJudge using the procedure described above. The final Replication Score is the average of the three scores, mitigating outliers and providing a more stable and reliable measure of the model's typical performance.

\begin{figure}[b]
    \centering
    \includegraphics[width=\textwidth]{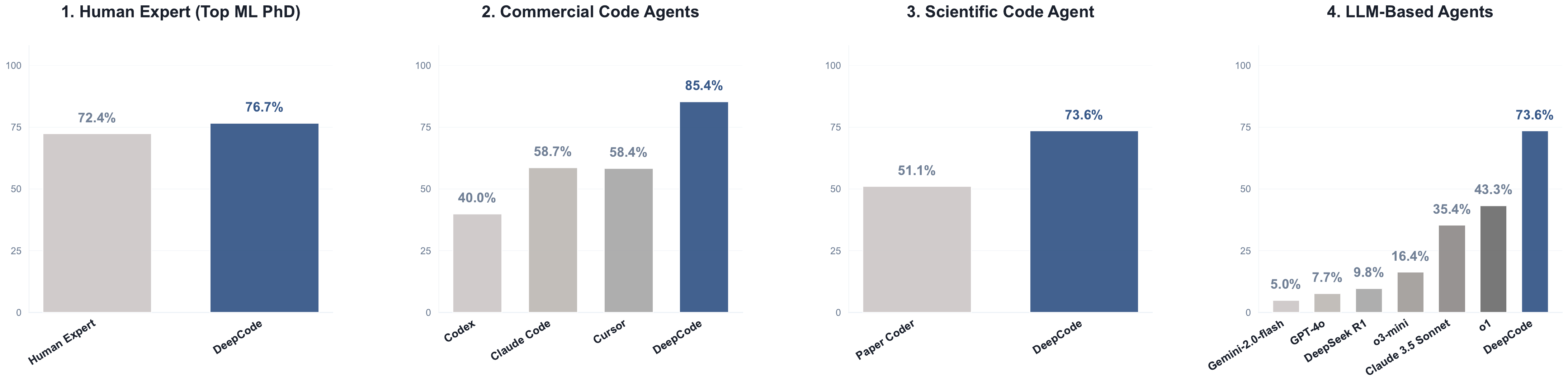}
    \caption{Comparison of \model\ with four baseline categories: (1) human experts, (2) state-of-the-art commercial code agents, (3) scientific code agents, and (4) LLM-based agents}
    \label{fig:main_result}
\end{figure}

\subsection{Main Results}
The primary results of our experiments are detailed in Figure~\ref{fig:main_result}. We analyze the performance of \model\ against the four established categories of baselines: general-purpose LLM agents, specialized scientific code agents, commercial code agents, and human experts.

\begin{itemize}[leftmargin=*]
\item \textbf{Comparison against LLM Agents.} Figure~\ref{fig:main_result} presents average replication scores across all benchmark papers. Among general-purpose LLM agents, performance varies significantly by model and scaffolding. With BasicAgent, Claude-3.5-Sonnet achieves the highest score (35.4$\bf{\pm}$0.8), while other frontier models range from 5.0 to 19.5. IterativeAgent scaffolding improves some models, with o1 reaching the best LLM agent performance of 43.3$\bf{\pm}$1.1. \model\ achieves 73.5$\bf{\pm}$2.8, representing a 70\% relative improvement over the best LLM agent baseline. This substantial gap demonstrates that our framework's specialized design, which incorporates systematic planning, structured code generation and  automated verification, provides significant advantages over general-purpose agent scaffolding.

\item \textbf{Comparison against Scientific Code Agents.} PaperCoder, a specialized multi-agent framework designed for transforming machine learning papers into executable code, achieves a score of 51.1$\bf{\pm}$1.4, outperforming all LLM agents baselines. However, \model\ achieves a significantly higher score of 73.5$\bf{\pm}$2.8—an improvement of over 22 points. This substantial gain suggests that our approach to task decomposition, code generation, and repository-level integration is markedly more effective than existing specialized methods.

\item \textbf{Comparison against Commercial Code Agents.} Table~\ref{tab:main2} details a direct comparison with leading commercial agents on a 5-paper subset. \model\ achieves an average score of 0.8482, decisively outperforming Codex (0.3997), Cursor (0.5841), and Claude Code (0.5871).
This result is particularly noteworthy: \model\ uses the same base model as both Cursor and Claude Code. The dramatic performance difference provides strong evidence that our framework's performance gains are not merely a product of a powerful base model. Rather, the advantage is directly attributable to the superior agentic architecture, planning, and execution strategies of \model.

\item \textbf{Comparison against Human Experts.} The most compelling finding is the comparison to human expert performance. As shown in the final rows of Figure~\ref{fig:main_result}, we benchmarked performance on the 3-paper subset. The human baseline, which represents the best-of-3 attempts from ML PhD students, achieved a score of 72.4. Our \model's average performance on this same subset was 75.9 $\bf{\pm}$ 4.5, meaning it not only competes with but exceeds the score of the best attempt from a human expert. This result strongly validates our approach, demonstrating its capability to automate and even surpass expert-level performance on this highly challenging task.
\end{itemize}

\begin{table}[h]
\renewcommand\arraystretch{1.3}
\centering
\caption{Reproduction scores of \model\ and commercial code agents on 5-paper subset}
\scalebox{0.96}{
\begin{tabular}{lcccccc}
\hline
\textbf{Model} & \textbf{fre} & \textbf{rice} & \textbf{bam} & \textbf{pinn} & \textbf{mech-u} & \textbf{Avg.} \\
\hline
Codex (GPT 5 Codex-high) & 0.4095 & 0.3645 & 0.1937 & 0.5382 & 0.4926 & 0.3997 \\
Claude Code (Claude Sonnet 4.5-think) & 0.6286 & 0.3787 & 0.3829 & 0.7233 & 0.8222 & 0.5871 \\
Cursor (Claude Sonnet 4.5-think) & 0.6344 & 0.4186 & 0.3779 & 0.7748 & 0.7148 & 0.5841 \\
\hline
\textbf{\model} (Claude Sonnet 4.5-think) & \textbf{0.8435} & \textbf{0.7380} & \textbf{0.8530} & \textbf{0.9474} & \textbf{0.8888} & \textbf{0.8541} \\
\hline
\end{tabular}
}
\label{tab:main2}
\end{table}

\subsection{Analysis on Different LLMs} 

We evaluate \model\ with five LLM backbones (Claude-4.5-Sonnet, GPT-5, Claude-3.5-Sonnet, Gemini-2.5-Pro, DeepSeek-R1) on three PaperBench tasks (fre, all-in-one, stay-on-topic). The tasks vary in specification complexity: fre and all-in-one contain long, interdependent setups with overlapping constraints, while stay-on-topic provides more structured descriptions. Agent architecture and tooling remain constant to isolate model capability effects.

\begin{figure}[ht]
    \centering
    \includegraphics[width=0.9\textwidth]{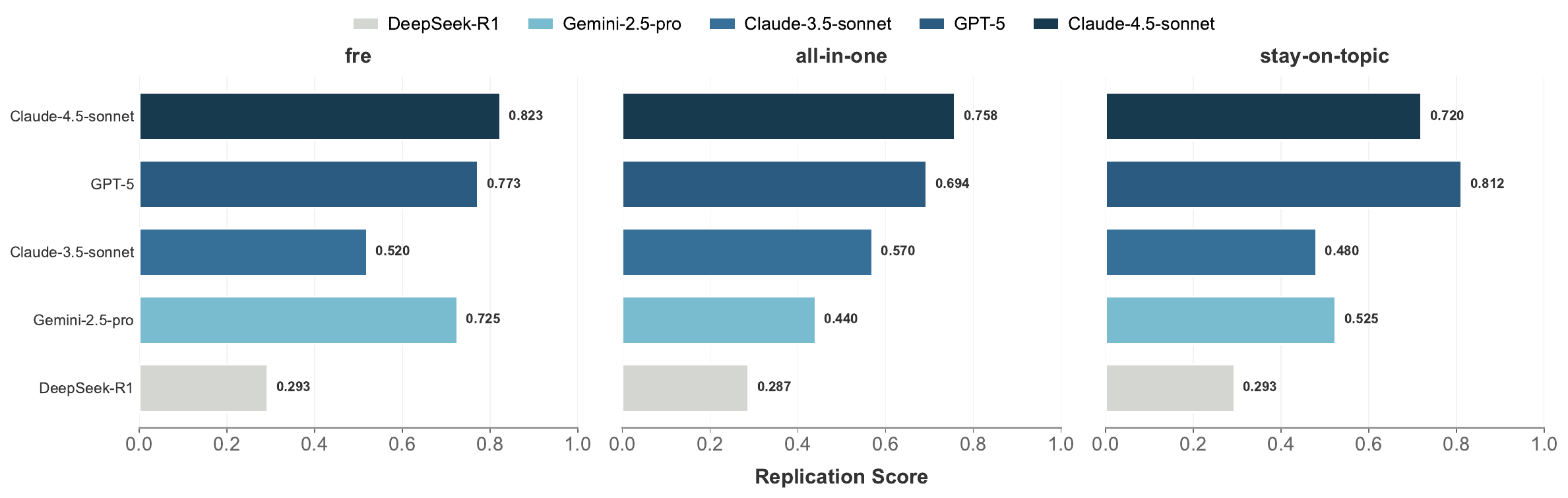}
    \caption{\model\ reproduction results on the 3-paper subset across LLM backbones}
    \label{fig:different_model_result}
    \vspace{-0.15in}
\end{figure}

As shown in Figure~\ref{fig:different_model_result}, reproduction scores exhibit consistent stratification across all three tasks. Claude-4.5-Sonnet achieves the best or near-best performance (0.72-0.82), demonstrating particular strength on fre and all-in-one where it more reliably reconstructs implementation details and multi-stage pipelines implied by complex, underspecified descriptions. GPT-5 tracks Claude-4.5-Sonnet closely on most metrics (0.69-0.81) and shows marginal advantages on stay-on-topic (0.81 vs. 0.72), suggesting additional robustness in maintaining alignment with fixed experimental framings, though this does not overturn Claude-4.5-Sonnet's overall dominance. Mid-tier models occupy an intermediate performance range: Claude-3.5-Sonnet (0.48-0.57) and Gemini-2.5-Pro (0.44-0.73) successfully recover main experimental skeletons but leave notable gaps in finer-grained procedural steps. DeepSeek-R1 consistently underperforms ($\approx$0.29), reproducing only fragments of target workflows across all tasks. This stable ranking pattern across heterogeneous specifications indicates that under fixed agent architecture, the underlying language model becomes the primary factor determining the ceiling and reliability of automatic paper-level reproduction.

\subsection{Ablation Studies}

In this section, we conduct ablation studies on three core components of \model: CodeRAG, CodeMem, and Automated Verification. Specifically, we evaluate CodeRAG and Automated Verification on a 3-paper subset (all-in-one, fre, stay-on-topic), while CodeMem is assessed on 5 randomly selected tasks (test-time-model-adaptation, rice, mechanistic-understanding, fre, all-in-one). Our key findings are summarized as follows.

\begin{figure}[b]
    \vspace{-0.2in}
    \centering
    \begin{subfigure}[t]{0.47\textwidth}
        \centering
        \includegraphics[width=1.1\textwidth]{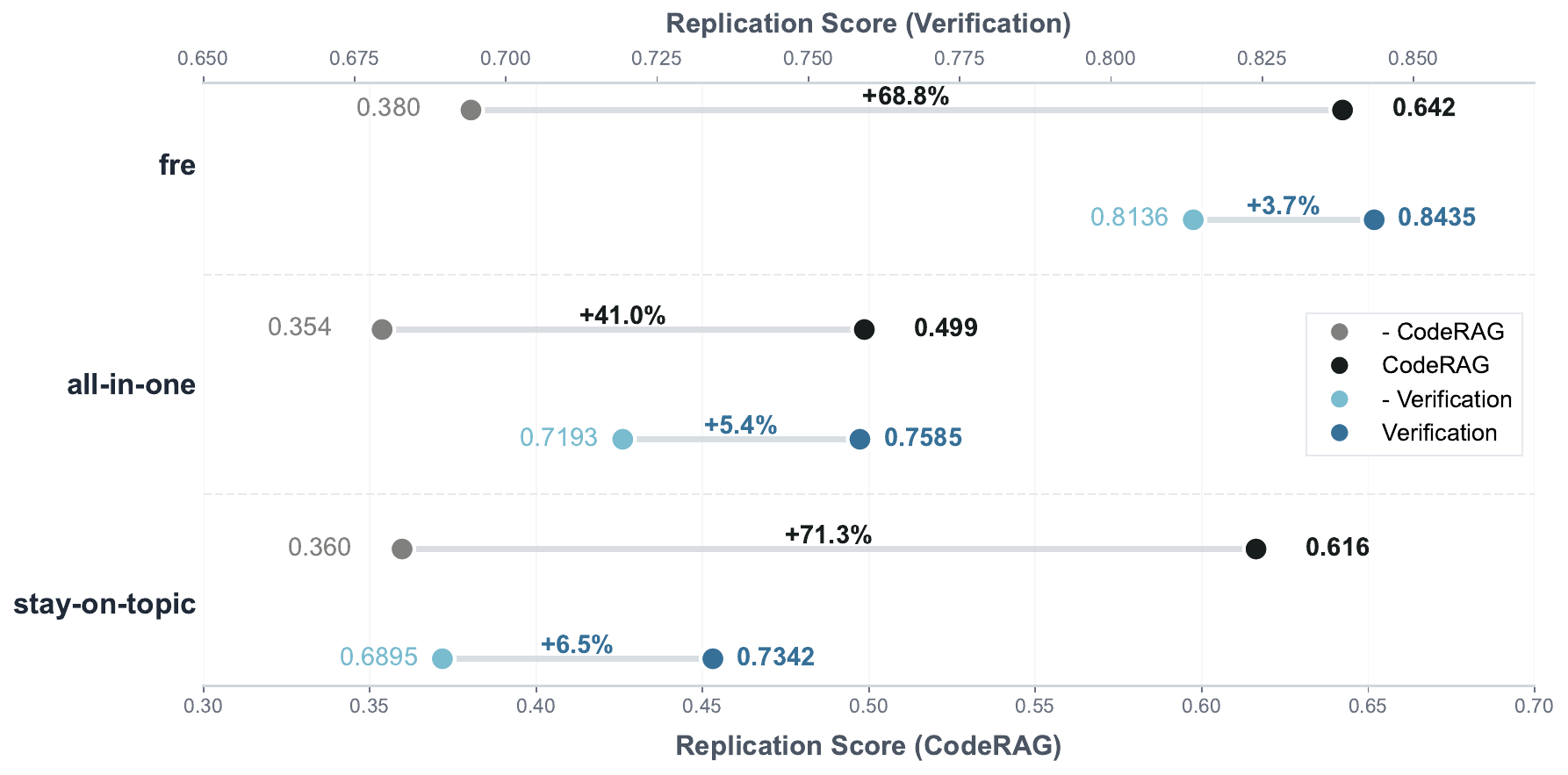}
        \caption{Ablation of CodeRAG and Verification}
        \label{fig:ablation_rag}
    \end{subfigure}
    \begin{subfigure}[t]{0.42\textwidth}
        \centering
        \includegraphics[width=0.8\textwidth]{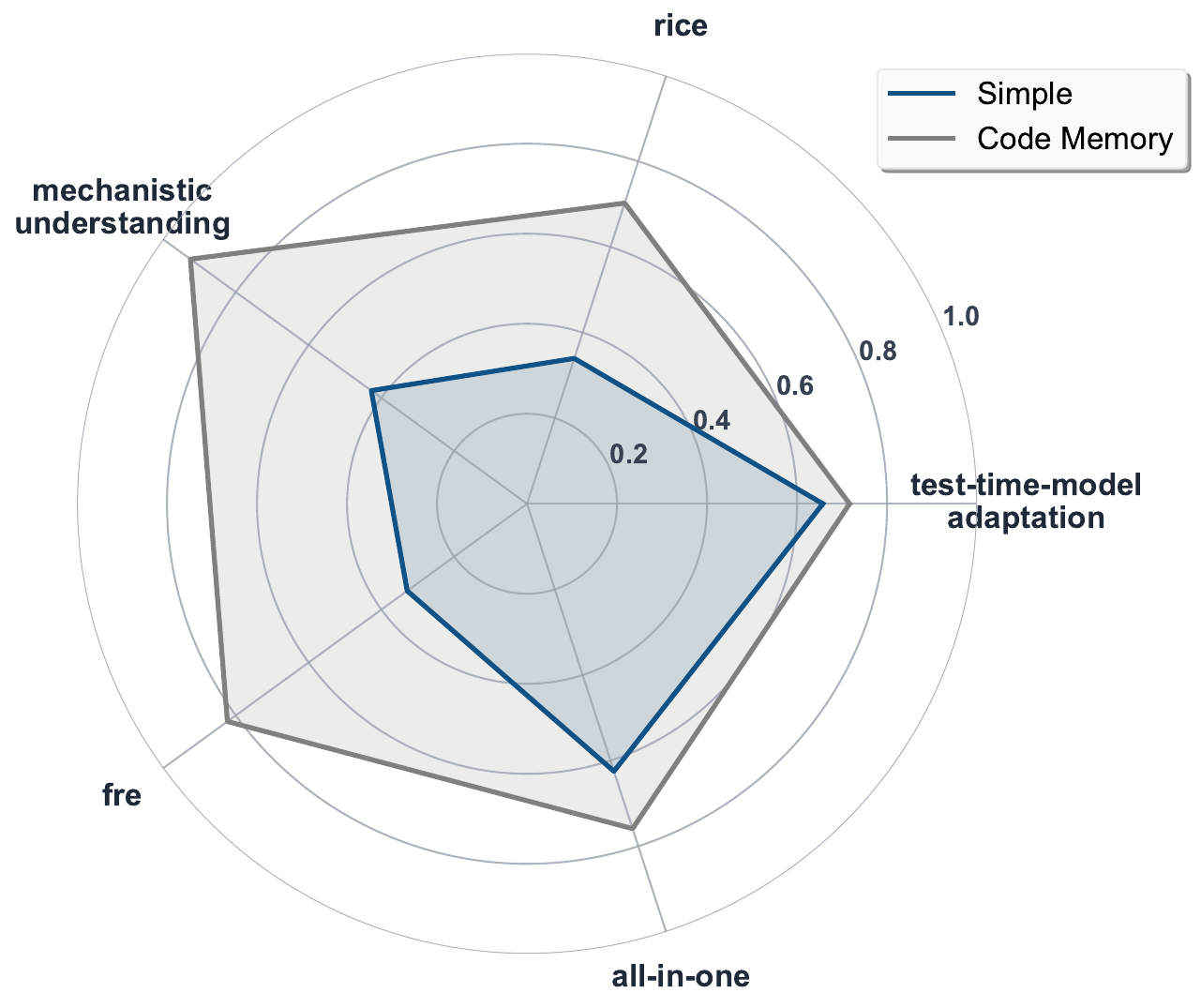}
        \caption{Ablation of CodeMem}
        \label{fig:ablation_memory}
    \end{subfigure}
 
    \caption{Ablation studies of key components in \model\ on PaperBench}
    \label{fig:ablation_overall}
\end{figure}

\textbf{(1) Impact of CodeRAG.} To decouple the impact of CodeRAG, we conducted an ablation study using Gemini-2.5-Flash. As visualized in Figure \ref{fig:ablation_rag}, the integration of CodeRAG delivers a substantial performance leap (up to 70\% relative gain), effectively breaking the base model's performance ceiling (0.35–0.38). Notably, we observed negligible gains when applying CodeRAG to frontier models like Claude 4.5 Sonnet. This contrast yields a critical insight: while reasoning giants likely encode sufficient implementation patterns within their parameters, cost-efficient models like Flash suffer from inherent \emph{knowledge gaps}. Consequently, CodeRAG proves indispensable for these architectures, acting as a vital bridge to fill implicit domain voids with standard practices—confirming that external knowledge injection is essential for democratizing high-fidelity replication on lightweight models.

\textbf{(2) Impact of CodeMem.} 
We ablate CodeMem's contribution on five PaperBench tasks using Claude-4.5-Sonnet, comparing \model's structured memory against a "Simple" baseline that naively evicts historical messages via sliding windows when approaching context limits.

\vspace{-0.05in}
Results demonstrate that unstructured eviction causes context saturation with signal loss: the Simple protocol achieves only 0.33-0.43 in rice, fre, and mechanistic-understanding tasks due to dependency truncation, where foundational class definitions are discarded before dependent code generation. CodeMem's structured indexing maintains task-relevant signal density, restoring scores to 0.70-0.92 by preserving critical dependencies without exhausting context budgets.
Even in scenarios with strong baseline performance (test-time-model-adaptation: 0.62 → 0.72; all-in-one: 0.66 → 0.76), Structured memory delivers consistent gains, confirming our core thesis: effective agentic coding requires explicit information flow management to maximize signal-to-noise ratio under context constraints.

\textbf{(3) Impact of Automated Verification.}
Across 3 test papers, Automated Verification yields consistent gains of 3.7–6.5\%, elevating scores from 0.69–0.81 to 0.73–0.84. The layer primarily corrects three types of residual errors: typos in variable names, missing dependencies, and wrong command-line arguments. These errors prevent otherwise sound implementations from executing reliably. The modest improvement reflects an important fact: the earlier phases have already achieved technical correctness. Verification is a final pass to ensure reliable execution. It eliminates small but consequential deviations that cause borderline implementations to fail, transforming them into faithful replications.


\section{Related Work}
\label{sec:related_work}

\subsection{General Coding Agents}

The field of software engineering is being rapidly transformed by agentic systems that have evolved from passive code assistants into autonomous entities capable of planning, executing multi-step tasks, and self-correction~\cite{dong2025survey,ge2025survey}. Research has explored several key architectures for these agents. One prominent trend involves multi-agent frameworks that emulate human development teams. This includes systems like ChatDev~\cite{qian2024chatdev}, MetaGPT~\cite{hong2024MetaGPT}, and CodePoRi~\cite{rasheed2024codepori}, which simulate entire software company organizational structures to manage development tasks from scratch. 
For repo-level code generation, CodeS~\cite{zan2024codes} proposed to decompose repository generation into specialized agents for structure planning and content filling. AgentCoder~\cite{huang2024agentcoder} employs atest-driven refinement loop involving programmer, test designer, and test executor agents, while MapCoder~\cite{islam2024mapcoder} mirrors human program synthesis with four agents handling example retrieval, planning, generation, and debugging.
A second major trend focuses on enhancing agents with specialized tools and interfaces. For instance, CodeAgent~\cite{zhang2024codeagent} integrates five domain-specific tools to support repository-level analysis, while SWE-agent~\cite{yang2025sweagent} introduces a high-level Agent-Computer Interface (ACI) to enable robust agent interaction with file systems and development environments. In addition, ToolGen~\cite{wang2025toolgen} proposes representing each tool as a unique token and directly integrating tool-specific knowledge into the parameters of the LLM, thereby enabling a paradigm shift toward seamless unification of tool invocation and natural language generation.

Recent advancements in academic research are increasingly being translated into practical, productized tools. Commercial code agents emerging from this trend can be broadly categorized into two distinct paradigms: (1) AI-native integrated development environments (IDEs) such as Cursor~\cite{cursor2025} and Trae~\cite{trae2025} that embed AI capabilities directly into the editor interface, and (2) terminal-based or extension-based agents including Claude Code~\cite{claudecode2025}, Gemini CLI~\cite{geminicli2025}, Github Copilot~\cite{copilot2025}, and Cline~\cite{cline2024} that operate through command-line interfaces or editor extensions. 
These coding agents leverage a holistic understanding of the codebase to perform complex tasks such as multi-file refactoring and autonomous edits. They support flexible, composable workflows and integrate seamlessly into diverse development pipelines. Commercial deployments indicate significant improvements in both function implementation and overall programming productivity.
Despite their effectiveness, these agents suffer from context window limitations that impair their ability to process lengthy technical documents such as academic papers, and struggle to maintain coherence and correctness when synthesizing repository-level codebases.

\subsection{Scientific Coding Agents}

In contrast to general-purpose coding agents, this class of agents targets more complex code generation scenarios, including the implementation and reproduction of entire codebases from high-level ideas and academic papers.
For example, Paper2Code~\cite{seo2025paper2code} addresses the research reproducibility crisis by transforming machine learning papers into executable repositories. Its code generation framework follows a structured three-stage process that includes system architecture design, implementation detail extraction, and modular code generation. CodeScientist~\cite{jansen2025codescientist} generates experimental code from literature, employing an iterative generate-execute-reflect cycle to write, run, and debug Python experiments. In addition, AlphaEvolve~\cite{novikov2025alphaevolve} utilize code generation for algorithmic discovery, using an LLM as an evolutionary mutator to propose variations to entire codebases, which are then rigorously evaluated.
Besides, the automation code in AI Scientist~\cite{lu2024aiscientist} and AI-Researcher~\cite{tang2025airesearcher} enables agents to iteratively plan and execute experiments, handle errors, and refine future runs based on results. AI Scientist focuses on experimental automation, maintaining execution history and generating plots and notes to support scientific write-ups. AI-Researcher extends this with a multi-stage refinement framework, where a code agent implements modular solutions and an advisor agent provides structured feedback for iterative validation, revision, and scaling.
These agents have advanced the pace of scientific research, yet achieving higher generation efficiency without compromising code quality remains an open challenge.
\section{Discussion: Challenges and Future Directions}
\label{sec:discussion}

While \model\ demonstrates the efficacy of principled information-flow management in high-fidelity repository synthesis, the transition from episodic coding tasks to autonomous, cost-effective, and self-evolving engineering remains fraught with challenges. We identify three critical frontiers that define the future trajectory of agentic software engineering.

\textbf{(1) Agentic Capability and Computational Efficiency. }
SOTA performance in agentic coding currently relies on massive, proprietary LLMs (\eg~GPT-5, Claude 4.5), which incur prohibitive deployment costs and high latency. Conversely, smaller, open-weight models offer efficiency but lack the complex reasoning capabilities required for autonomous decision-making in open-ended engineering tasks. 
Bridging this gap presents a dichotomy of challenges. \emph{(i) Fine-tuning limits:} Enhancing small models via supervised fine-tuning (SFT) is constrained by a data bottleneck—while raw code is abundant, high-quality agentic trajectories are scarce and expensive to curate. \emph{(ii) Knowledge injection limits:} Merely augmenting small models with external knowledge is often insufficient; retrieved contexts may lack direct relevance to the specific coding task, and small models struggle to integrate complex inputs without suffering from attention dilution.

We envision a shift toward hybrid agentic architectures that synergize models of varying scales, employing large models for high-level reasoning and efficient small models for routine implementation. Besides, distilling knowledge from large models helps reduce the data bottleneck.

\textbf{(2) From Episodic to Evolving Agents.}
Current coding agents typically operate in an episodic manner: they reset after each project, failing to carry over experience or tacit knowledge to subsequent tasks. Enabling agents to self-evolve and accumulate expertise mirrors human professional growth but faces significant hurdles.
\emph{(i) Reinforcement Learning constraints:} While RL-based optimization theoretically allows agents to learn from feedback, it requires well-defined reward functions, which are difficult to formulate for complex, multi-objective software engineering tasks. Moreover, this approach is inapplicable to closed-source LLMs where parameter updates are impossible. \emph{(ii) Memory scalability issues:} The alternative approach—stacking historical experiences into a long-term memory—introduces severe noise. Simply accumulating raw interaction logs leads to context bloat, where retrieving relevant past experiences becomes a ``needle in a haystack'' problem.

Beyond relying on extensive manual annotation and training, a scalable solution involves automating the abstraction of past experiences. Future agents can implement post-task reflection to condense execution traces into reusable skills or heuristics. Storing these refined insights allows agents to retrieve corresponding high-level guidance, enabling self-evolution while avoiding context explosion.

\textbf{(3) Dynamic Planning and Adaptability.}
Most existing frameworks utilize a linear Plan-then-Code workflow, assuming that all constraints are knowable a priori. In real-world engineering, specifications often evolve, and critical implementation constraints are frequently discovered only during the coding process.
Separation between planning and execution leads to fragility: if the initial blueprint is flawed, the coding agent is often constrained by a stale plan, leading to suboptimal workarounds or failure.

Future researches advance toward dynamic, bidirectional planning frameworks. Agents are able to adapt their initial blueprints when encountering unforeseen constraints during implementation. Establishing a feedback mechanism where execution insights directly inform and update the high-level plan is crucial for handling the complex realities of large-scale software development.

\vspace{-0.05in}
\section{Conclusion}
\vspace{-0.05in}
In this work, we presented \model, an autonomous framework that advances the frontier of agentic code engineering by reimagining document-to-repository synthesis as a challenge of \emph{information-flow management}. 
Addressing the fundamental conflict between information overload and finite context bottlenecks, we demonstrated that treating synthesis as a channel optimization problem—solved through the orchestration of blueprint distillation, stateful memory, conditional knowledge injection, and closed-loop verification—effectively maximizes the signal-to-noise ratio for long-horizon tasks. 
Empirical evaluations on PaperBench confirm that \model\ establishes a new SOTA, decisively outperforming leading commercial agents and surpassing PhD-level human experts in reproduction accuracy. 
These findings validate that hierarchical information orchestration, rather than indiscriminate context scaling, provides the decisive path toward robust autonomous systems, laying a critical foundation for the future of automated scientific discovery and rigorous research reproduction.

\clearpage

\bibliographystyle{unsrtnat}
\bibliography{neurips_2024}
\clearpage
\appendix

\section{Appendix}
This appendix supplements the main text by providing four categories of supplementary materials.
First, the \emph{Complete Results} subsection reports an extensive quantitative evaluation of \model,
including comparative analysis against multiple benchmark models and reproducibility analysis across different papers and operational scenarios.
Second, the \emph{DeepCode Application Cases} subsection showcases representative visualizations demonstrating DeepCode's end-to-end capabilities, covering backend systems, web user interfaces, and the Paper2Code research reproduction workflow.
Third, the \emph{DeepCode Sub-Agent Details} subsection elucidates the internal multi-agent architecture, clarifying the roles, responsibilities, and coordination patterns for implementing specific specialized sub-agents.
Finally, the \emph{MCP Toolkit in DeepCode} subsection documents the Model Context Protocol (MCP) tools integrated into the system, defining the external interfaces through which DeepCode interacts with code repositories, documentation, and execution environments.
\subsection{Full Results}
This appendix reports quantitative results that complement the main text and provide a more systematic evaluation of \model’s overall capability and stability on research code reproduction tasks. Table~\ref{tab:main1} first compares, under a unified evaluation protocol, a range of general-purpose code execution agents (including both BasicAgent and IterativeAgent configurations), existing specialized reproduction systems such as PaperCoder, and human experts on the same benchmark. \model\ achieves an average reproduction score of $73.5 \pm 2.8$ on the full benchmark, substantially outperforming PaperCoder ($51.1 \pm 1.4$) as well as all configurations derived from commercial models. On the 3-paper subset, \model\ attains an average score of $75.9 \pm 4.5$, exceeding the human “Best@3” score of 72.4, indicating that, on representative deep learning papers, the system delivers reproduction quality comparable to or better than that of strong human practitioners.

Table~\ref{tab:main2} further selects a 5-paper subset (fre, rice, bam, pinn, mech-u) for a head-to-head comparison against several widely used commercial code assistants (Codex, Claude Code, Cursor, etc.). Across all papers, \model\ achieves the highest reproduction score, with an average of 0.8482, corresponding to an absolute improvement of more than 0.26 over the strongest competing system. The advantage is consistent across all individual papers, suggesting that the gains arise from architectural and procedural design choices rather than from favorable alignment with a narrow subset of tasks.

Finally, Table~\ref{tab:papers_results} provides per-paper details for the Claude 4.5 Sonnet–based configuration, including three independent runs, their mean and standard error, as well as the associated average cost. Across a diverse set of targets—such as FRE, PINN, MECHANISTIC-UNDERSTANDING, and SEQUENTIAL-NEURAL-SCORE-ESTIMATION—\model’s reproduction scores typically lie in the 0.7–0.9 range with relatively small standard errors, while the distribution of average cost across papers remains tight. This indicates strong cross-task generalization, stable behavior across repeated runs, and reasonable resource usage. Taken together, these appendix results reinforce the main conclusions of the paper: on realistic research code reproduction benchmarks, \model\ not only achieves significantly higher average performance than existing automated reproduction and code assistance systems, but also demonstrates robust and consistent advantages in fine-grained, multi-paper, multi-run analyses.
\begin{table}[h]
\renewcommand\arraystretch{1.2}
\centering
\caption{Average reproduction scores: \model\ vs. LLMs and human experts}
\begin{tabular}{lr}
\hline
\textbf{Model} & \textbf{Average Replication Scores} \\
\hline
GEMINI-2.0-FLASH (BasicAgent) & $5.0 \pm 0.0$ \\
4o (BasicAgent) & $7.7 \pm 0.0$ \\
o3-mini (BasicAgent) & $5.1 \pm 0.8$ \\
o1 (BasicAgent) & $19.5 \pm 1.2$ \\
R1 (BasicAgent) & $9.8 \pm 0.0$ \\
CLAUDE-3-5-SONNET (BasicAgent) & $35.4 \pm 0.8$ \\
o3-mini (IterativeAgent) & $16.4 \pm 1.4$ \\
o1 (IterativeAgent) & $43.3 \pm 1.1$ \\
CLAUDE-3-5-SONNET (IterativeAgent) & $27.5 \pm 1.6$ \\
o1 [36 hours] (IterativeAgent) & $42.4 \pm 1.0$ \\
\hline
PaperCoder & $51.1 \pm 1.4$  \\
\textbf{\model} & \textbf{73.6} $\bf{\pm}$ \textbf{5.3}  \\
\hline
\hline
\textbf{Human} [3 paper subset, Best@3] & 72.4 \\
\textbf{\model} [3 paper subset, Average] & \textbf{76.7} $\bf{\pm}$ \textbf{3.9}\\
\hline
\hline
\end{tabular}
\label{tab:main1}
\end{table}

\begin{table}[htbp]
    \centering
    \caption{\model\ with Claude 4.5 Sonnet results.}
    \label{tab:papers_results}
    \resizebox{\textwidth}{!}{%
    \setlength{\tabcolsep}{4pt}
    \renewcommand{\arraystretch}{1.1}
    \begin{tabular}{@{}l *{6}{c}@{}}
        \toprule
        \textbf{Paper} & \textbf{Run 1} & \textbf{Run 2} & \textbf{Run 3} & \textbf{Mean} & \textbf{Std. Error} & \textbf{Avg. Cost} \\
        \midrule
        FRE & 0.844 & 0.823 & 0.803 & 0.814 & 0.020 & 9.14 \\
        RICE & 0.738 & 0.609 & 0.761 & 0.702 & 0.082 & 8.22\\
        BAM & 0.853 & 0.673 & 0.719 & 0.748 & 0.094 & 8.45\\
        WILL-MODEL-FORGET & 0.776 & 0.793 & 0.857 & 0.808 & 0.042 & 9.20\\
        PINN & 0.947 & 0.800 & 0.983 & 0.910 & 0.097 & 7.84\\
        ALL-IN-ONE & 0.769 & 0.747 & 0.759 & 0.759 & 0.011 & 9.43\\
        ADAPTIVE-PRUNING & 0.547 & 0.570 & 0.516 & 0.544 & 0.027 & 9.13\\
        LBCS & 0.689 & 0.732 & 0.820 & 0.747 & 0.066 & 10.01\\
        MECHANISTIC-UNDERSTANDING & 0.889 & 0.944 & 0.941 & 0.925 & 0.031 & 10.20\\
        TEST-TIME-MODEL-ADAPTATION & 0.717 & 0.578 & 0.652 & 0.649 & 0.069 & 7.90\\
        SAMPLE-SPECIFIC-MASKS & 0.690 & 0.740 & 0.583 & 0.671 & 0.080 & 8.30\\
        BRIDGING-DATA-GAPS & 0.552 & 0.566 & 0.626 & 0.581 & 0.039 & 7.98\\
        STAY-ON-TOPIC-WITH-CLASSIFIER-FREE-GUIDANCE & 0.734 & 0.800 & 0.626 & 0.705 & 0.088 & 9.12\\
        STOCHASTIC-INTERPOLANTS & 0.851 & 0.792 & 0.801 & 0.815 & 0.031 & 8.89\\
        LCA-ON-THE-LINE & 0.665 & 0.844 & 0.739 & 0.749 & 0.090 & 7.73\\
        SEQUENTIAL-NEURAL-SCORE-ESTIMATION & 0.930 & 0.862 & 0.817 & 0.870 & 0.057 & 10.01\\
        SAPG & 0.702 & 0.755 & 0.757 & 0.738 & 0.031 & 9.19\\
        FTRL & 0.558 & 0.606 & 0.631 & 0.598 & 0.037 & 7.06\\
        ROBUST-CLIP & 0.772 & 0.742 & 0.685 & 0.733 & 0.044 & 7.83\\
        BBOX & 0.620 & 0.681 & 0.631 & 0.644 & 0.033 & 11.90\\
        \bottomrule
    \end{tabular}%
    }
\end{table}
\subsection{Use Cases for DeepCode}
This appendix provides a series of visual artifacts generated by DeepCode, offering concrete evidence of its capabilities across different software development and research domains. These examples are intended to supplement the main paper by illustrating the practical utility and versatility of our system.

\begin{figure}[t]
    \centering
    \includegraphics[width=0.48\textwidth]{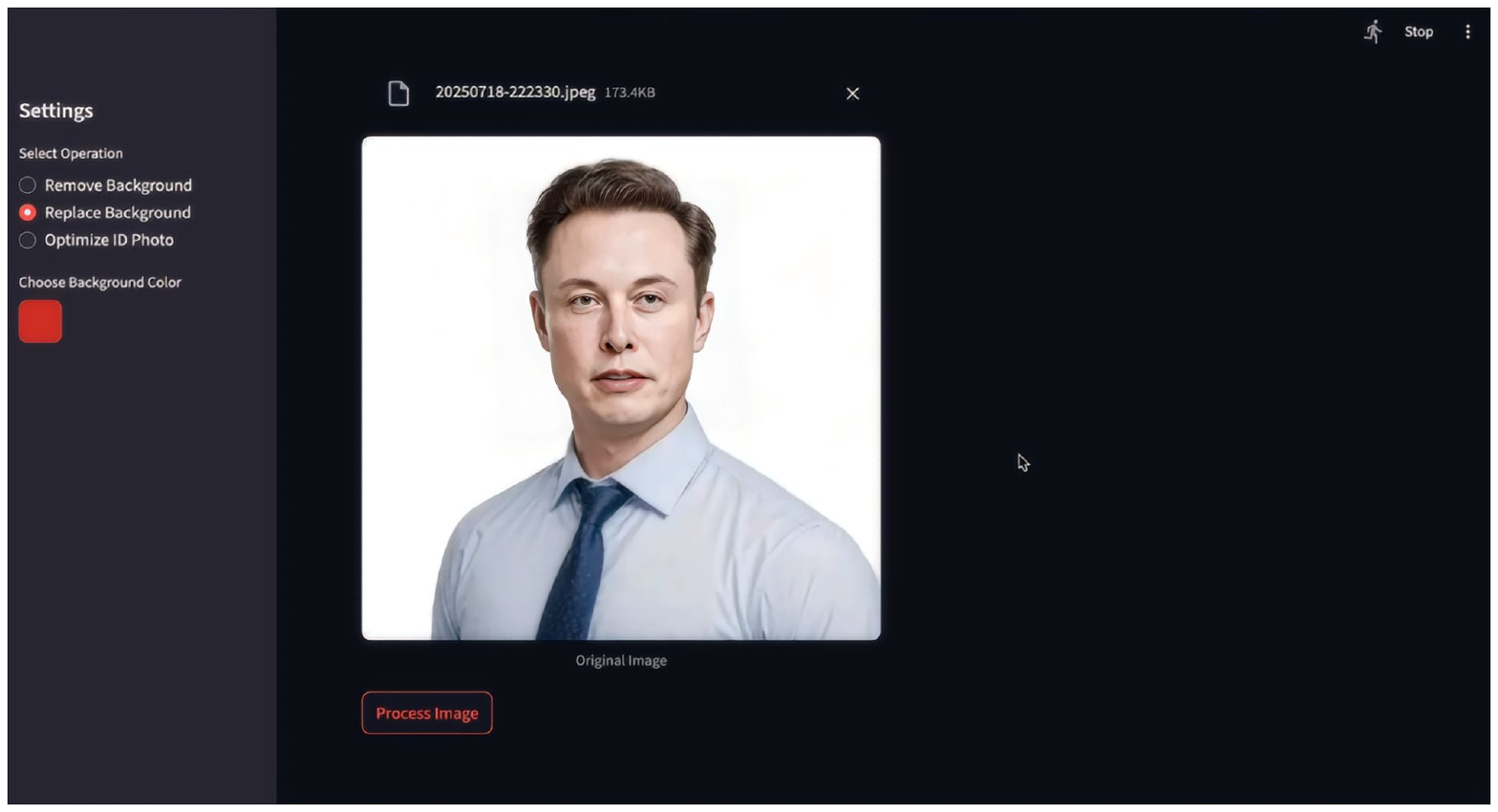}
    \includegraphics[width=0.48\textwidth]{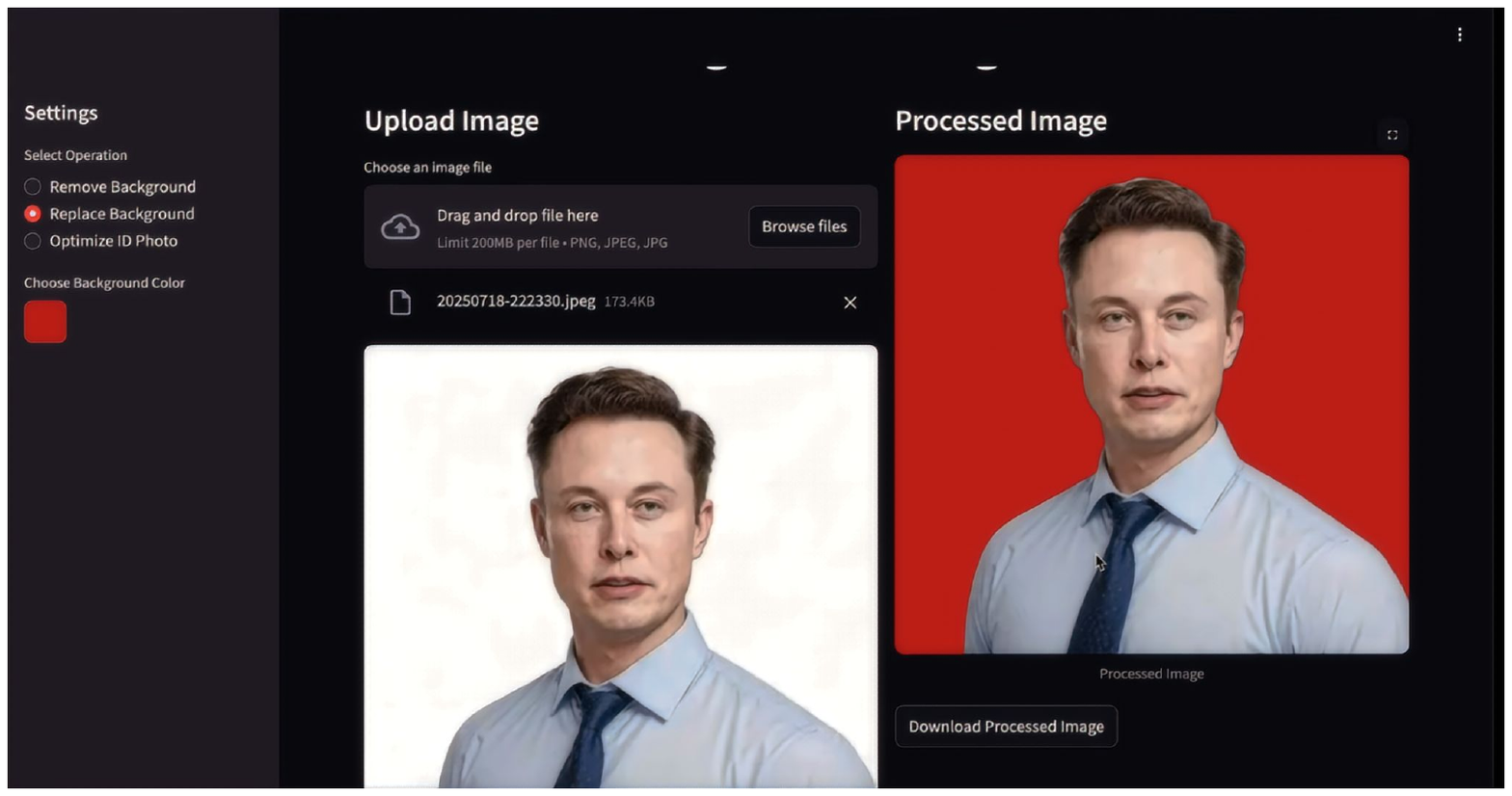}
    \caption{DeepCode-generated backend system pages.}
    \label{fig:deepcode_backend_pages}
\end{figure}

The initial set of examples, depicted in Figure \ref{fig:deepcode_backend_pages}, focuses on DeepCode's proficiency in generating sophisticated backend systems. The figures showcase automatically constructed administrative dashboards, which likely include functionalities for data monitoring, user management, and content moderation. Such pages are critical for the operational management of modern web applications but are often tedious and repetitive to build. DeepCode's ability to scaffold these complex, data-driven interfaces from high-level specifications demonstrates its potential to significantly reduce boilerplate engineering and accelerate the deployment of robust server-side infrastructure.

\begin{figure}[t]
    \centering
    \includegraphics[width=0.48\textwidth]{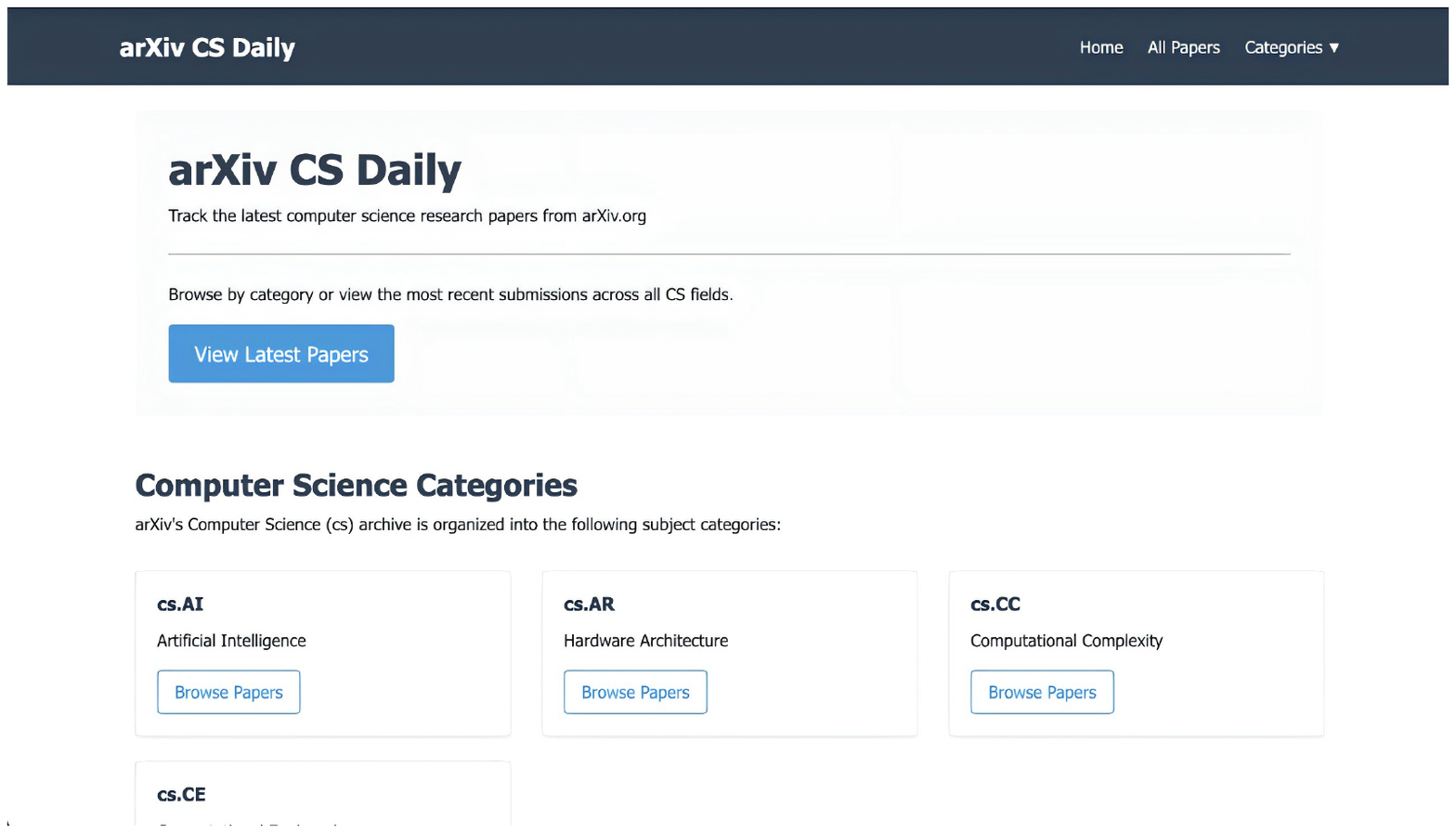}
    \includegraphics[width=0.48\textwidth]{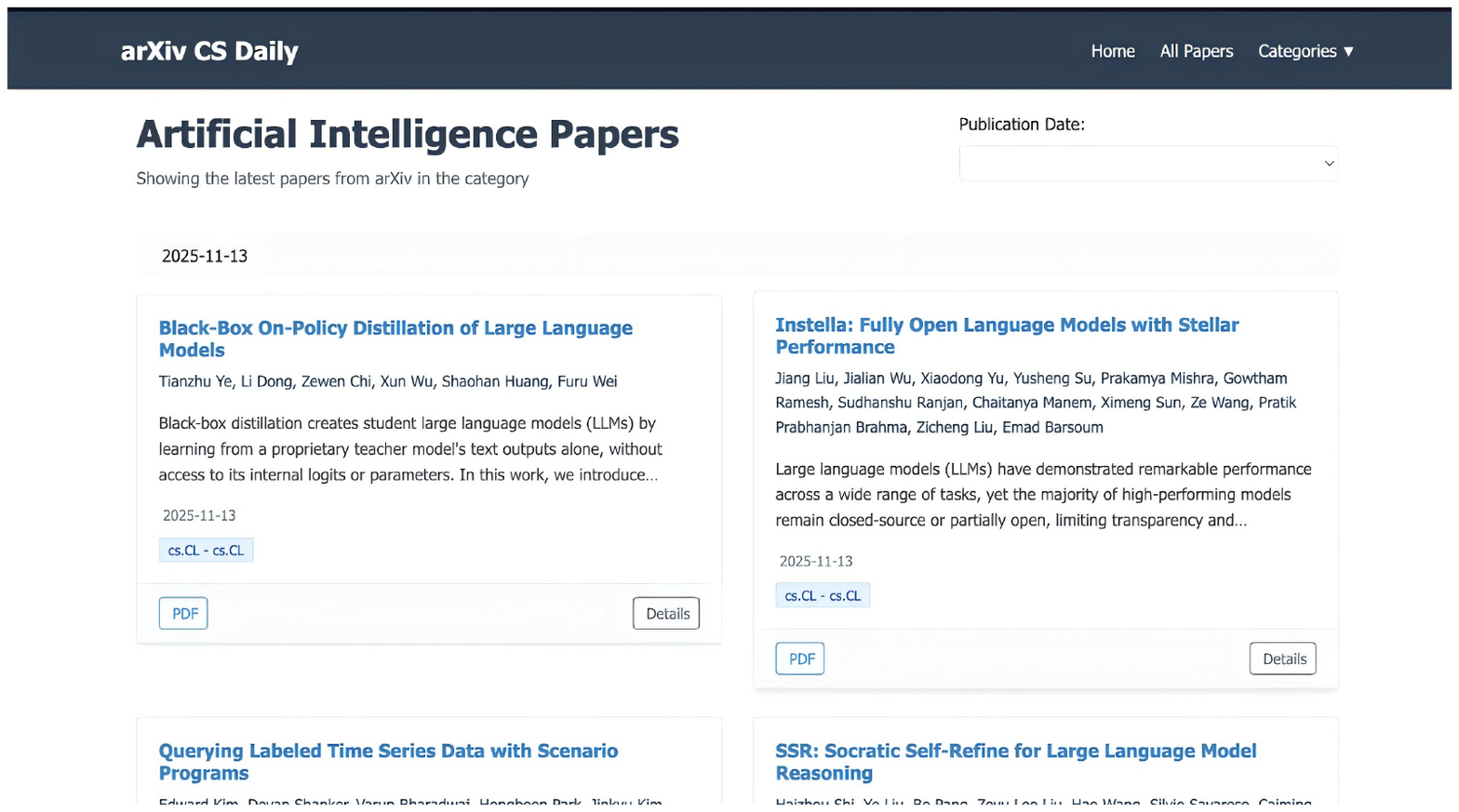}
    \caption{DeepCode-generated Web UI.}
    \label{fig:deepcode_web}
\end{figure}

Building upon the backend logic, a system's utility is often defined by its user-facing presentation. Figure \ref{fig:deepcode_web} illustrates DeepCode's capacity for generating intuitive and functional Web UIs. The generated interfaces, featuring elements such as data visualization charts and interactive forms, translate abstract user requirements into tangible, interactive components. This capability not only complements the backend generation by providing a corresponding frontend, but also empowers developers and designers to rapidly prototype and iterate on user experiences, thereby shortening the path from concept to a functional product.

\begin{figure}[t]
    \centering
        \includegraphics[width=0.19\textwidth]{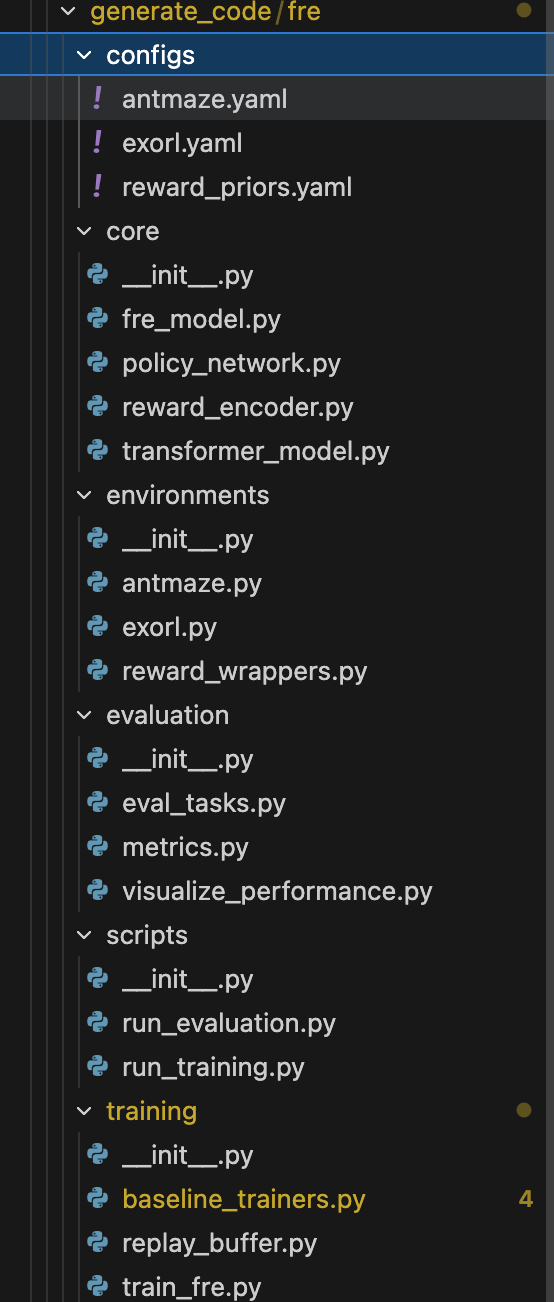}
        \includegraphics[width=0.7\textwidth]{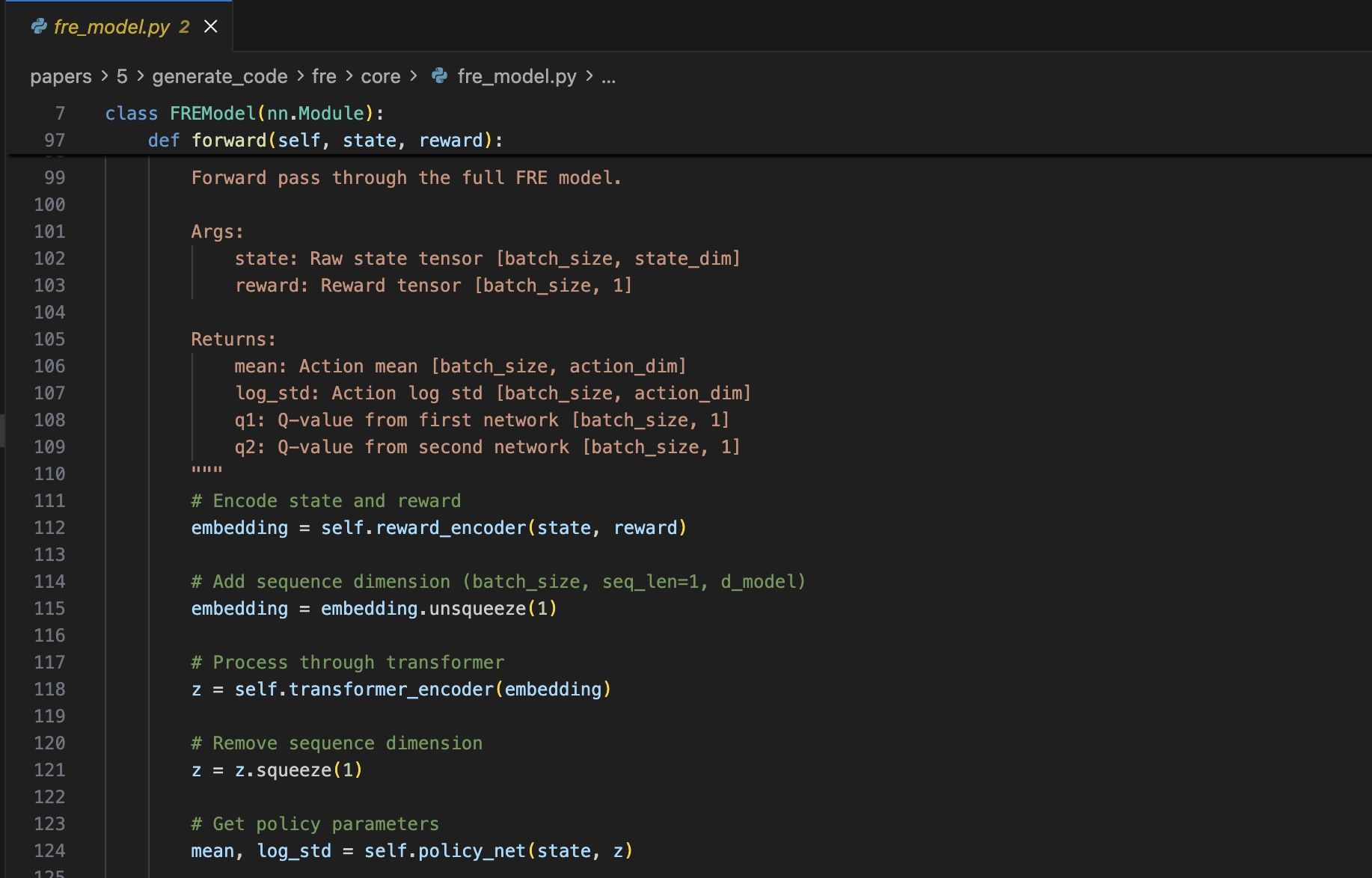}

    \caption{Paper2Code Samples of DeepCode. Left: Code Structure, Right: Code Sample}
    \label{fig:paper2code_overall}
\end{figure}

Perhaps DeepCode's most ambitious application, however, lies in its potential to bridge the chasm between academic research and practical implementation. The Paper2Code functionality, illustrated in Figure \ref{fig:paper2code_overall}, exemplifies this capability. The figure is twofold: on the left, it presents the high-level code structure that DeepCode inferred from a research paper, discerning the architectural blueprint of the proposed algorithm, including its modular components and file organization. On the right, it provides a concrete code sample, instantiating a specific function or class with precise logic. This powerful feature moves beyond conventional code generation by interpreting unstructured scientific language to produce structured, executable artifacts, thereby holding immense promise for enhancing research reproducibility and accelerating the adoption of novel scientific discoveries.
\subsection{Sub-Agents Details of DeepCode}
DeepCode decomposes the software engineering pipeline into a set of specialized
agents with narrow, well-specified responsibilities and standardized
communication interfaces, rather than relying on a single monolithic generative
model. The individual agents and their responsibilities are summarized in
Table~\ref{tab:deepcode_agents}. This modular design allows different stages of
the lifecycle—ranging from requirement understanding to architectural planning
and code synthesis—to be implemented as transformations over shared
intermediate representations, while preserving global architectural and
semantic consistency.

\textbf{During the planning stage, DeepCode relies on explicit coordination
between conceptual and algorithmic analysis agents to derive a coherent
development blueprint from high-level specifications.}
The Central Orchestrating Agent first routes each input through the Document
Parsing and/or Intent Understanding agents to obtain a structured
specification, which then serves as the input to the Code Planning agent.
Within this planning module, two internal analysis pipelines operate in
parallel over the same intermediate representation. The conceptual analysis
sub-agent is responsible for system-level decomposition: it identifies major
subsystems, their responsibilities, and inter-module interfaces, and it
constructs an architecture-level call topology. The algorithmic analysis
sub-agent is responsible for computational aspects: it abstracts key
algorithmic ideas, selects candidate data structures, reasons about time and
space complexity constraints, and enumerates feasible implementation patterns.
The partial plans produced by these two sub-agents are reconciled by a
planning aggregation component (Code Analysis agent), which resolves
inconsistencies and materializes a project-level development roadmap, including
module boundaries, interface signatures, dependency relations, implementation
priorities, and testing hooks. This roadmap serves as the design baseline that
constrains all downstream code generation and refinement steps.

\textbf{During the code synthesis stage, DeepCode couples retrieval-augmented
reference mining with a global code memory, forming a closed-loop process that
enforces repository-level consistency during incremental generation.}
On the retrieval side, the Code Reference Mining and Code Indexing agents
implement a Retrieval-Augmented Generation (RAG) layer: they maintain
multi-granularity indices over a corpus of prior implementations and expose to
the Code Generation agent semantically relevant and structurally compatible
code patterns, ranging from individual functions to reusable design idioms. In
parallel, the Code Memory agent maintains a structured representation of the
current repository state, including cross-file symbol tables, dependency
graphs, and project-wide conventions such as naming schemes, error-handling
strategies, and configuration mechanisms. Before emitting new code, the Code
Generation agent issues queries to the Code Memory agent to obtain the
up-to-date repository context and applicable constraints; after generation, it
writes back the newly introduced symbols and dependencies, triggering an update
of the global repository model. This query–constraint–update loop allows
DeepCode to align local synthesis decisions with global architectural intent,
reducing interface mismatches, naming drift, and latent coupling across the
codebase.
\begin{table}[htbp]
    \centering
    \caption{Functional Specifications of Specialized Sub-Agents in the DeepCode Framework}
    \label{tab:deepcode_agents}
    \renewcommand{\arraystretch}{1.4}
    \begin{tabular}{p{0.28\linewidth} p{0.67\linewidth}}
        \toprule
        \textbf{Agent Role} & \textbf{Functional Description} \\
        \midrule
        \textbf{Central Orchestrating Agent} & 
        Functions as the central control unit, responsible for task decomposition, resource allocation, and the strategic coordination of sub-agents based on the complexity of the input requirements. \\
        
        \textbf{Intent Understanding Agent} & 
        Conducts semantic parsing of natural language inputs to extract functional requirements, converting ambiguous user descriptions into formal technical specifications. \\
        
        \textbf{Document Parsing Agent} & 
        Processes unstructured technical documents (e.g., research papers). It extracts multimodal information, including text, mathematical formulas, and diagrams, to establish a ground truth for implementation. \\
        
        \textbf{Concept Analysis Agent} & 
        Abstracts core theoretical concepts and logical flows from the parsed specifications, ensuring the computational model aligns with the theoretical underpinnings of the source material. \\
        
        \textbf{Algorithm Analysis Agent} & 
        Evaluates and selects appropriate algorithmic strategies and data structures. It focuses on optimizing computational complexity and feasibility before code synthesis begins. \\
        
        \textbf{Code Planning Agent} & 
        Formulates the software architecture and development roadmap. This agent determines the technology stack, designs modular file structures, and enforces design patterns to ensure scalability. \\
        
        \textbf{Code Reference Mining Agent} & 
        Retrieves external knowledge by identifying relevant open-source repositories. It analyzes dependency graphs to recommend integration patterns and library usages. \\
        
        \textbf{Code Memory Agent} & 
        Manages the state and context throughout the generation lifecycle. It utilizes hierarchical data structures to retain historical decisions and maintain semantic consistency across long-context interactions. \\
        
        \textbf{Code Generation Agent} & 
        Synthesizes executable source code based on the architectural plan and retrieved references. It implements functional interfaces and integrates distinct modules into a cohesive codebase. \\
        
        \textbf{Automated Validation Agent} & 
        Executes a rigorous quality assurance loop. It performs static analysis, generates unit tests, and iteratively debugs the codebase to verify functional correctness and adherence to specifications. \\
        \bottomrule
    \end{tabular}
\end{table}
\subsection{MCP Tool Stack in DeepCode}
\label{appendix:mcp-tools-subsec}

Table~\ref{tab:deepcode_mcp_tools} summarizes the Model Context Protocol (MCP) tools integrated into DeepCode. The tools are grouped into three functional categories: \emph{Perception \& Retrieval}, \emph{Cognitive Processing}, and \emph{Action \& Execution}. This organization makes the main stages of the system explicit. Perception \& Retrieval tools give the model access to up-to-date web search results, web pages, and binary documents such as research papers and technical manuals, which helps mitigate the effects of the model’s knowledge cut-off. Cognitive Processing tools then convert large codebases and long documents into semantic indexes and context-window-compatible segments, so that the model can issue natural language queries over existing artifacts and work with long technical materials. Action \& Execution tools finally operate on the local development environment by reading and writing project files, executing shell commands, and interacting with the version control system.

Taken together, the tools in Table~\ref{tab:deepcode_mcp_tools} form an end-to-end loop for assisted software development. The system can retrieve external and local information, reorganize it into internal structures that fit within the model’s context window, and then apply code changes while observing their effects through commands such as tests or package installations. The table also shows that operations with side effects on the environment (file I/O, command execution, and Git operations) are confined to the \emph{Action \& Execution} layer and are described as sandboxed and path-validated. This separation between information access, semantic processing, and environment manipulation makes the extension of the base language model through MCP tools transparent and easier to reason about.
\begin{table}[h!]
    \centering
    \caption{Specification of Model Context Protocol (MCP) Tools Integrated into DeepCode. These tools extend the Large Language Model's capabilities across perception, cognitive processing, and environment manipulation domains}
    \small
    \renewcommand{\arraystretch}{1.5}
    \begin{tabularx}{\textwidth}{|l|l|X|}
    \hline
    \textbf{Category} & \textbf{MCP Server Name} & \textbf{Functional Description \& Academic Specification} \\
    \hline
    \multirow{8}{*}{\textbf{Perception \& Retrieval}} 
    & \texttt{brave\_search} & 
    A real-time information retrieval interface leveraging the Brave Search API. It provides the agent with temporal-aware access to web indices, enabling the retrieval of up-to-date documentation and resolving knowledge cut-off limitations. \\
    \cline{2-3}
    & \texttt{bocha\_mcp} & 
    A specialized search module delivering structured "modal cards" and semantic summaries. It serves as a secondary knowledge source, optimizing token efficiency by returning structured entities rather than raw HTML. \\
    \cline{2-3}
    & \texttt{fetch} & 
    A web content ingestion engine that retrieves URL endpoints and normalizes heterogeneous HTML structures into clean Markdown. It acts as the agent's primary reading interface for external documentation. \\
    \cline{2-3}
    & \texttt{pdf\_downloader} & 
    Binary resource acquisition tool designed for academic papers and technical manuals. It handles HTTP streams to ingest non-textual document formats (PDF/DOCX) for downstream processing. \\
    \hline
    \multirow{6}{*}{\textbf{Cognitive Processing}} 
    & \texttt{code\_reference\_indexer} & 
    A Retrieval-Augmented Generation (RAG) module for local codebases. It constructs a vector or semantic index of the project files, allowing the agent to perform natural language queries over the existing code structure. \\
    \cline{2-3}
    & \texttt{document\_segmentation} & 
    A pre-processing utility implementing semantic chunking algorithms. It partitions large technical documents into context-window-compliant segments, facilitating the "Paper2Code" workflow for complex algorithm implementation. \\
    \hline
    \multirow{10}{*}{\textbf{Action \& Execution}} 
    & \texttt{filesystem} & 
    A sandboxed file I/O interface allowing controlled read/write operations within the project directory. It enforces path validation security policies to prevent unauthorized system access during code generation. \\
    \cline{2-3}
    & \texttt{code\_implementation} & 
    The core generative engine encapsulated as an MCP tool. It orchestrates the synthesis of functional code blocks, integrating logic planning with atomic file writing operations to ensure code coherence. \\
    \cline{2-3}
    & \texttt{command\_executor} & 
    A runtime environment interface permitting the execution of shell commands (e.g., \texttt{pytest}, \texttt{pip install}). It establishes a feedback loop by capturing \texttt{stdout}/\texttt{stderr} for iterative debugging and self-correction. \\
    \cline{2-3}
    & \texttt{git\_command} & 
    Version control management interface. It abstracts Git plumbing commands, enabling the agent to manage repository state, branch for experimental features, and maintain a clean commit history. \\
    \hline
    \end{tabularx}
    \label{tab:deepcode_mcp_tools}
\end{table}

\end{document}